\documentclass[preprint]{ptptex} 

\usepackage{wrapft}
\usepackage{amsmath,amssymb,amscd,amsbsy,amsgen,amsopn,amstext,amsxtra}
\usepackage[mathscr]{eucal}

\title{Spinor and Twistor Formulations of Tensionless \\ 
Bosonic Strings in Four Dimensions}

\author{Shinichi \textsc{Deguchi},$^{1,}$\footnote{E-mail:  deguchi@phys.cst.nihon-u.ac.jp} 
Takeshi \textsc{Egami}$^{2,}$\footnote{E-mail:  egami@phys.cst.nihon-u.ac.jp} 
and Jun-ichi \textsc{Note}$^{2,}$\footnote{E-mail:  note@phys.cst.nihon-u.ac.jp} 
}

\inst{
$^1$Institute of Quantum Science, College of Science and Technology, \\ 
Nihon University, Tokyo 101-8308, Japan
\\
$^2$Graduate School of Quantum Science and Technology, \\ 
Nihon University, Tokyo 101-8308, Japan
}
\abst{%
Spinor and twistor formulations of tensionless bosonic strings  
in 4-dimensional Minkowski space are constructed. 
We begin with a first-order action that is equivalent to the Nambu-Goto action 
in the tensionful case and that leads to a spinorial action in the tensionless case. 
From this spinorial action, we find an alternative 
spinorial action useful for constructing 
a simple twistor formulation of tensionless strings. 
The twistor formulation is steadily constructed  
in accordance with a fundamental concept of twistor theory. 
We investigate local internal symmetries inherent in 
the twistorial action for a tensionless string and carry out 
some classical analyses of the tensionless string expressed in a twistorial form. 
}

\begin{document} 
\maketitle

\numberwithin{equation}{section}

\section{Introduction}

Tensionless strings, or null strings, have been studied from various 
aspects \cite{Sch, KL, LRSS, AB, GRR, ILST, DL, Bon},  
since Schild gave an earlier formulation of tensionless bosonic strings \cite{Sch}. 
As mentioned in the literature, tensionless strings are expected to be 
realized as the extremely high-energy limit of ordinary tensionful strings. 
Also, tensionless strings might be useful to illustrate a configuration of QCD strings 
that occurs in the transition from the confinement phase to 
the deconfinement phase. 
In many of the local field theories, particle masses are treated as 
secondary contents provided by mass-generation mechanisms   
such as the spontaneous symmetry breaking.  
If this view can be extended to string theory,  
the string tension would be a secondary part provided  
by some unknown mechanism. 
In this context, we should begin with tensionless strings, 
and then should consider a mechanism for deriving tensionful strings. 
Interesting ideas for generating string tension have actually been proposed in 
Ref.~\citen{Tow}.

To quantize strings, it is necessary to define the ordering prescription 
for the Hamiltonian operator of a string \cite{GSW-Kak}. 
It is known that tensionless strings can be consistently quantized not only with  
normal ordering but also with Weyl ordering.  
The normal ordering prescription leads to the existence of critical dimensions,  
as in the quantization of ordinary tensionful strings:  
the quantization with normal ordering is consistent only  
at $D=26$ for tensionless bosonic strings  
and at $D=10$ for tensionless supersymmetric strings \cite{GRR}.  
In contrast, the Weyl ordering prescription causes no  
critical dimensions \cite{LRSS, AB, GRR}. 
This makes us expect that tensionless strings in 4-dimensions can consistently 
be formulated at the quantum-theoretical level as well as at the classical level.

Here, we turn our attention to the fact 
that the twistor theory provides elegant descriptions of 
4-dimensional systems having no mass scales \cite{PM, Pen2, PR, Pen, HT, Tak}.   
Using twistor variables, 
Shirafuji has constructed a simple Lagrangian formalism 
for massless spinning particles in 4-dimensional Minkowski space \cite{Shi}.  
This was steadily accomplished through rewriting 
an ordinary action for a massless particle in terms of the space-time coordinates  
and the 4-momentum represented as a bilinear form in 2-component spinors.  
A similar spinor representation of the momentum-like bivector was introduced 
by Gusev and Zheltukhin to define a 2-component spinor formulation  
of tensionless bosonic strings in 4-dimensional Minkowski space \cite{GZ}.  
The action proposed by them is certainly a spinorial action 
for a tensionless string. 
However, in their approach, it is unclear how this action is derived from a vector formulation 
of tensionless strings, in comparison with the particle case. 
Although Gusev and Zheltukhin's action governs a spinor formulation of tensionless strings,  
it is difficult to rewrite this action only in terms of twistor variables. 
Hence, Gusev and Zheltukhin's action is not appropriate for constructing a {\em genuine} 
twistor formulation of tensionless strings. 
A twistor formulation of tensionless bosonic strings has actually been proposed by Ilyenko \cite{Ily}. 
However, in this approach, we cannot expect a simple canonical treatment, 
because Ilyenko's action for a tensionless string consists of terms quartic in two null twistors 
and their dual twistors.

The purpose of this paper is to construct a simple genuine twistor formulation of 
tensionless bosonic strings in 4-dimensional Minkowski space 
and to carry out some classical analyses in this formulation. 
To steadily achieve the purpose, we begin with a first-order formalism for bosonic strings 
that leads, in the tensionful case, to the Nambu-Goto string theory \cite{NG}. 
In the tensionless case, this formalism gives a pair of simultaneous constraints   
that can be solved in terms of 2-component spinors. 
With the spinor solution to the pair of constraints, the first-order action of a tensionless bosonic string 
is written in a spinorial form. 
The spinorial action obtained here is essentially identical to Gusev and Zheltukhin's action; 
however, we derive it in a systematic way on the basis of the first-order formalism. 
We also carry out a classical analysis based on the spinorial action 
to confirm the relation to the first-order formalism.  
For convenience,  
next we find an alternative spinorial action from Gusev and Zheltukhin's action. 
Also, we verify that the alternative spinorial action yields Gusev and Zheltukhin's action. 
Although an action similar to the alternative spinorial action is seen in Ref.~\citen{BZ}, 
its construction is essentially different from ours. 
Our approach makes it clear that the alternative spinorial action actually 
describes a tensionless bosonic string.

We construct a genuine twistor formulation of tensionless strings by considering   
the spinor formulation governed by the alternative spinorial action. 
This is performed in accordance with a concept of twistor theory that 
twistors are more primitive than the space-time coordinates \cite{PM, PR, Pen}. 
We therefore start from an action consisting of a twistor, its dual twistor 
and Lagrange multiplier fields. 
Then, we show that this action reduces to the alternative spinorial action 
by imposing the null twistor condition. 
It is also shown that the null twistor condition can naturally be obtained by 
modifying the action so as to leave the modified one invariant under 
the local phase transformation of the twistors. 
The modified action remains invariant not only under the local phase transformation  
but also under the local scale transformation of the twistors.  
For this reason, the twistors constituting the modified action are treated  
as {\em projective} twistors. 
We carry out some classical analyses on the basis of 
the modified action written in terms of rescaled twistors and Lagrange multiplier fields. 
This action yields two constraint conditions: the null twistor condition and 
a subsidiary condition associated with reparametrizations of the string spatial parameter. 
With the Dirac brackets defined for the rescaled twistors, it is demonstrated that 
the two constraints constitute a closed algebra involving the (classical) Virasoro algebra 
without central extension. 
We also mention a resemblance between our twistor formulation of tensionless strings 
and the twistor formulation of massive particles \cite{Pen2, Per, Hug}.

The present paper is organized as follows: 
Section 2 presents a first-order formalism for bosonic strings in 4-dimensional 
Minkowski space and examines 
reparametrization symmetry of a tensionless string. 
In \S3, the first-order action of a tensionless string, given in \S2,  
is rewritten in terms of 2-component spinors.  
Here, it is pointed out that the resulting spinorial action is essentially identical to   
Gusev and Zheltukhin's action. 
A simpler equivalent form of the spinorial action is found in \S4. 
Because this form is not fully reparametrization invariant, it is improved to 
an alternative spinorial action in such a way that the full reparametrization invariance 
is restored.  
In \S5, we construct a twistor formulation of tensionless bosonic strings, 
verifying that it can reduce to the spinor formulation treated in \S4. 
Section 6 considers local phase and scale symmetries in the twistor formulation.  
In \S7, we carry out some classical analyses in the twistor formulation.  
Section 8 is devoted to a summary and discussion. 
Appendix A contains a summary of 2-component spinor notation 
and a solution to a pair of simultaneous constraint equations. 
Appendix B provides a twistor formulation of massless particles for comparison with 
our twistor formulation of tensionless strings.

\section{A first-order formalism for bosonic strings}

We treat a string propagating in 4-dimensional Minkowski space $\mathbf{M}$ 
with the metric tensor $\eta_{\mu\nu}=\mathrm{diag}(1,-1,-1,-1)$. 
Let $x^{\mu}=x^{\mu}(\tau, \sigma)$ $\,(\mu=0, 1, 2, 3)$ be space-time coordinates of a point 
on the world sheet, $\varSigma$, that the string sweeps out as it moves in time.  
Here, $\tau$ $(\tau_{0} \leq \tau \leq \tau_{1})$ is a parameter describing time development of  
the string, and $\sigma$ $(\sigma_{0}\leq \sigma \leq \sigma_{1})$  
is a spatial coordinate along the string. 
Then, it follows that 
$(\dot{x}^{\mu})$ [$\,\dot{x}^{\mu} := \partial x^{\mu}/\partial \tau\,$] 
is a time-like or null vector, 
while $(\acute{x}^{\mu})$ [\,$\acute{x}^{\mu} := \partial x^{\mu}/\partial \sigma$\,]  
is a spacelike vector:  
\begin{subequations}
\label{2.1}
\begin{align}
\dot{x}^{2} &:= \dot{x}_{\mu} \dot{x}^{\mu} \geq 0 \,, 
\label{2.1a}
\\
\acute{x}^{2} &:= \acute{x}_{\mu} \acute{x}^{\mu} < 0 \,.
\label{2.1b}
\end{align}
\end{subequations}
The world sheet $\varSigma$ is defined by the mapping  
$(\tau, \sigma) \mapsto x^{\mu}(\tau, \sigma)$ 
from the 2-dimensional parameter space 
$\varXi:=\{(\tau, \sigma)\}$, a subspace of $\Bbb{R}^2$, to 
$\mathbf{M}$, so that $\varSigma$ is parametrized by $(\tau, \sigma)$. 
With this mapping, it is possible to identify $\varXi$ with $\varSigma$. 
The surface element of $\varSigma$ is given by $v^{\mu\nu} d\tau d\sigma$ with  
$v^{\mu\nu} := \dot{x}^{\mu} \acute{x}^{\nu}- \dot{x}^{\nu} \acute{x}^{\mu}$.

Let us consider the action 
\begin{align}
\hat{S}_{1}=\int_{\varXi} d\tau d\sigma \hat{\mathcal{L}}_{1} 
\label{2.2}
\end{align}
with the Lagrangian (density)\footnote{The Lagrangian (\ref{2.3}) 
remains invariant under the parity transformation 
$\big(x^{0}, x^{i}\big) \rightarrow \big(x^{0}, -x^{i} \big)$ $(i=1,2,3)$, 
provided that the relevant fields transform as  
$p_{0i} \rightarrow -p_{0i}$, $p_{ij} \rightarrow p_{ij}$, $e \rightarrow e$, and 
$f \rightarrow -f$. } 
\begin{align}
\hat{\mathcal{L}}_{1}=\frac{1}{2} p_{\mu\nu} v^{\mu\nu}
-\frac{1}{4} e p_{\mu\nu} p^{\mu\nu} -\frac{1}{4} f p_{\mu\nu} \tilde{p}^{\mu\nu} 
- \frac{e^2 +f^2}{2e} T^2 \,. 
\label{2.3}
\end{align}
Here, $T$ is a constant with the dimension of mass squared, 
$p_{\mu\nu}=p_{\mu\nu}(\tau,\sigma)$ 
are real scalar fields on $\varXi$ satisfying $p_{\mu\nu}=-p_{\nu\mu}$, 
and $\tilde{p}^{\mu\nu}$ denotes the Hodge dual components of $p_{\mu\nu}\,$: 
$\tilde{p}^{\mu\nu} :=\frac{1}{2} \epsilon^{\mu\nu\rho\sigma} p_{\rho\sigma}$ 
$\big(\epsilon^{0123}=-1\big)$.   
Also, $e=e(\tau, \sigma)$ and $f=f(\tau, \sigma)$ are real scalar-density fields 
on $\varXi$, that is,  real fields on $\varXi$ that 
$e d\tau d\sigma$ and $f d\tau d\sigma$ behave as scalars 
under the reparametrization 
\begin{align}
\tau \rightarrow \tau^{\prime}=\tau^{\prime} (\tau, \sigma)\,, 
\qquad 
\sigma  \rightarrow \sigma^{\prime}=\sigma^{\prime} (\tau, \sigma)\,. 
\label{2.4}
\end{align}
The space-time coordinates $x^{\mu}=x^{\mu}(\tau, \sigma)$ are regarded as   
real scalar fields on $\varXi$.  
Note that the action $\hat{S}_{1}$ is first order in $v^{\mu\nu}$. 
It is now evident that the action $\hat{S}_{1}$ remains invariant  
under the reparametrization (\ref{2.4}). 
Because the Lagrangian (\ref{2.3}) contains no derivative terms of $p_{\mu\nu}$,  
$e$ and $f$, these fields are treated as auxiliary fields.  
Variation of $\hat{S}_{1}$ with respect to $p_{\mu\nu}$ yields the equation 
\begin{align}
v_{\mu\nu}=e p_{\mu\nu} +f \tilde{p}_{\mu\nu} \,, 
\label{2.5}
\end{align}
whose Hodge dual is found to be 
\begin{align}
\tilde{v}_{\mu\nu}= -f p_{\mu\nu} +e \tilde{p}_{\mu\nu} \,, 
\label{2.6}
\end{align}
where $\tilde{v}_{\mu\nu} := \frac{1}{2} \epsilon_{\mu\nu\rho\sigma} v^{\rho\sigma}
=\epsilon_{\mu\nu\rho\sigma} \dot{x}^{\rho} \acute{x}^{\sigma}$.  
Equations (\ref{2.5}) and (\ref{2.6}) are inversely solved as 
\begin{align}
p_{\mu\nu}&=\frac{1}{e^2 +f^2} (ev_{\mu\nu} -f \tilde{v}_{\mu\nu}) \,, 
\label{2.7} 
\\ 
\tilde{p}_{\mu\nu}&=\frac{1}{e^2 +f^2} (f v_{\mu\nu} +e \tilde{v}_{\mu\nu}) \,. 
\label{2.8} 
\end{align}

Substituting Eqs. (\ref{2.7}) and (\ref{2.8}) into Eq. (\ref{2.3}) and using 
the identities $v_{\mu\nu} \tilde{v}^{\mu\nu}=0$ and 
$\tilde{v}_{\mu\nu} \tilde{v}^{\mu\nu}= -v_{\mu\nu} v^{\mu\nu}$, 
we obtain\footnote{
In a particular gauge $(e, f)=(1, 0)$, Eq. (\ref{2.9}) reduces to the Schild Lagrangian 
\cite{Sch}, up to an additional constant $-T^{2}/2$.}
\begin{align}
\hat{\mathcal{L}}_{1}=\frac{e}{4(e^2 +f^2)} \,v_{\mu\nu} v^{\mu\nu} 
- \frac{e^2 +f^2}{2e} T^2 \,. 
\label{2.9}
\end{align}
From this, the Euler-Lagrange equations for $e$ and $f$ are derived, 
respectively, as 
\begin{align}
\bigg( 1-\frac{f^2}{e^2} \bigg) K &=0 \,, 
\label{2.10}
\\ 
\frac{f}{e} K &=0\,, 
\label{2.11}
\end{align}
where 
\begin{align}
K :=\frac{e^2}{2(e^2 +f^2)^2} \,v_{\mu\nu} v^{\mu\nu} + T^2 \,.
\label{2.12}
\end{align}
Combining Eq. (\ref{2.10}) with Eq. (\ref{2.11}) leads to $K=0$, 
which can be written as 
\begin{align}
\frac{e^2 +f^2}{e}  T= \sqrt{ -\frac{1}{2}v_{\mu\nu} v^{\mu\nu} } 
\label{2.13}
\end{align}
under the condition $e>0$. 
In the case $T>0$, referred to as the tensionful case, 
it is possible to completely eliminate $e$ and $f$ from 
Eq. (\ref{2.9}) by using Eq. (\ref{2.13}). After eliminating $e$ and $f$, we have      
\begin{align}
\hat{\mathcal{L}}_{1}= -T \sqrt{ -\frac{1}{2}v_{\mu\nu} v^{\mu\nu} } \,. 
\label{2.14} 
\end{align}
The Lagrangian (\ref{2.14}) is precisely the Nambu-Goto Lagrangian with 
the tension parameter $T$ \cite{NG}. 
Thus, we see that in the tensionful case, the action $\hat{S}_{1}$ is classically equivalent  
to the Nambu-Goto action. The equivalence at the quantum-theoretical level can be 
demonstrated by means of the path-integral method.

Now, we return to the Lagrangian (\ref{2.3}).  From this, the Euler-Lagrange 
equations for $e$ and $f$ are derived, respectively, as   
\begin{align}
\frac{1}{2} p_{\mu\nu} p^{\mu\nu} + \bigg( 1-\frac{f^2}{e^2} \bigg) T^2 &=0 \,, 
\label{2.15} 
\\ 
\frac{1}{2} p_{\mu\nu} \tilde{p}^{\mu\nu} -\frac{2f}{e} T^2 &=0 \,. 
\label{2.16}
\end{align}
Substituting Eq. (\ref{2.7}) into Eq. (\ref{2.15}) yields Eq. (\ref{2.10}), and   
substituting Eqs. (\ref{2.7}) and (\ref{2.8}) into Eq. (\ref{2.16}) yields Eq. (\ref{2.11}). 
As shown above, Eq. (\ref{2.10}), together with Eq. (\ref{2.11}), leads to 
Eq. (\ref{2.13}).  In this way, we can directly derive Eq. (\ref{2.13}) from  
the Lagrangian (\ref{2.3}) without going through the Lagrangian (\ref{2.9}).

Next, let us consider the tensionless case, $T=0$, in which Eq. (\ref{2.2}) reads 
\begin{align}
S_{1}=\int_{\varXi} d\tau d\sigma \mathcal{L}_{1} 
\label{2.17}
\end{align}
with 
\begin{align}
\mathcal{L}_{1}=\frac{1}{2} p_{\mu\nu} v^{\mu\nu}
-\frac{1}{4} e p_{\mu\nu} p^{\mu\nu} -\frac{1}{4} f p_{\mu\nu} \tilde{p}^{\mu\nu} \,. 
\label{2.18}
\end{align}
Correspondingly, Eqs. (\ref{2.15}) and (\ref{2.16}) become 
\begin{align}
p_{\mu\nu} p^{\mu\nu} &=0 \,, 
\label{2.19}
\\
p_{\mu\nu} \tilde{p}^{\mu\nu} &=0 \,, 
 \label{2.20}
\end{align}
and Eq. (\ref{2.13}) reduces to  
\begin{align}
\frac{1}{2} v_{\mu\nu} v^{\mu\nu} 
=\dot{x}^{2} \acute{x}^{2} -(\dot{x} \cdot \acute{x})^{2}=0 \,, 
\label{2.21}
\end{align}
where $\dot{x} \cdot \acute{x} :=\dot{x}_{\mu} \acute{x}^{\mu}$. 
Equation (\ref{2.21}) is precisely the null world-sheet condition found by 
Schild \cite{Sch} at the initial stage of study of tensionless strings. 
Since the tensionless string is characterized by the condition  (\ref{2.21}), 
it is often refereed to as the null string.   
Equations (\ref{2.1a}) and (\ref{2.1b}) give $\dot{x}^{2} \acute{x}^{2} \leq 0$, 
whereas Eq. (\ref{2.21}) gives $\dot{x}^{2} \acute{x}^{2} \geq 0$ owing to   
$(\dot{x} \cdot \acute{x})^{2}\geq 0$. 
Both of these contrastive conditions are valid if and only if $\dot{x}^{2} \acute{x}^{2} =0$.  
Because $\acute{x}^{2} <0$, the condition $\dot{x}^{2} \acute{x}^{2} =0$ implies 
$\dot{x}^{2}=0$. Also, $\dot{x}^{2} \acute{x}^{2} =0$ leads to $\dot{x} \cdot \acute{x}=0$ 
via Eq. (\ref{2.21}). We thus obtain, from Eqs. (\ref{2.1}) and (\ref{2.21}),  
\begin{align}
\dot{x}^{2} &=0 \,, 
\label{2.22}
\\ 
\dot{x} \cdot \acute{x} &=0 \,. 
\label{2.23} 
\end{align}
Equations (\ref{2.22}) and (\ref{2.23}) imply together that each point on the string 
moves at the speed of light in a direction normal to the string curve at that point. 
In the ordinary (tensionful) string theory, Eq. (\ref{2.23}) is set by hand as a part of 
the so-called orthogonal gauge condition.  
By contrast, in the present tensionless model, 
Eq. (\ref{2.23}) is necessarily derived from the action $S_{1}$,   
supplemented with Eqs. (\ref{2.1}), as a (classical) condition inherent in the model.

Recall here that the space-time coordinates $x^{\mu}$ are scalar fields on $\varXi$, satisfying  
$x^{\prime\mu}(\tau^{\prime}, \sigma^{\prime})=x^{\mu}(\tau, \sigma)$. 
The derivatives of $x^{\prime\mu}$ 
with respect to $\tau^{\prime}$ and $\sigma^{\prime}$ are written as
\begin{align}
\dot{x}^{\prime\mu}=
\frac{\partial \tau}{\partial \tau^{\prime}} \dot{x}^{\mu} 
+\frac{\partial \sigma}{\partial \tau^{\prime}} \acute{x}^{\mu}\, , 
\qquad 
\acute{x}^{\prime\mu}=
\frac{\partial \tau}{\partial \sigma^{\prime}} \dot{x}^{\mu} 
+\frac{\partial \sigma}{\partial \sigma^{\prime}} \acute{x}^{\mu}\, .
\label{2.24}
\end{align}
Using Eqs. (\ref{2.22}), (\ref{2.23}) and (\ref{2.24}), we can show that 
\begin{align}
\dot{x}^{\prime\;\! 2} =
\left( \frac{\partial \sigma}{\partial \tau^{\prime}} \right)^{2} \acute{x}^{2} \,, 
\qquad 
\dot{x}^{\prime}\cdot\acute{x}^{\prime}=
\frac{\partial \sigma}{\partial \tau^{\prime}} \frac{\partial \sigma}{\partial \sigma^{\prime}}
\,\acute{x}^{2} \,,
\qquad 
\acute{x}^{\prime\;\! 2}=
\left( \frac{\partial \sigma}{\partial \sigma^{\prime}} \right)^{2} \acute{x}^{2} \,. 
\label{2.25}
\end{align}
The transformed coordinates $x^{\prime\mu}$ must satisfy the same properties as $x^{\mu}$, 
because there are no essential differences between $(\tau, \sigma)$ and 
$(\tau^{\prime}, \sigma^{\prime})$, and furthermore, the action $S_{1}$ 
is reparametrization invariant. 
Then, it follows that Eqs. (\ref{2.1}) and (\ref{2.21}) are valid also for 
$\dot{x}^{\prime\mu}$ and $\acute{x}^{\prime\mu}$, 
so that $\dot{x}^{\prime\!\; 2}\geq0$, $\acute{x}^{\prime\!\; 2}<0$, and 
$\dot{x}^{\prime\!\; 2} \acute{x}^{\prime\!\; 2} 
-(\dot{x}^{\prime} \cdot \acute{x}^{\prime}\:\!)^{2}=0$. 
Hence, following the procedure for deriving Eqs. (\ref{2.22}) and (\ref{2.23}), 
we obtain their counterparts 
$\dot{x}^{\prime\!\; 2}=0$ and $\dot{x}^{\prime}\cdot\acute{x}^{\prime}=0$. 
Taking into account Eq. (\ref{2.1b}),  
we see that the conditions $\dot{x}^{\prime\!\; 2}=
\dot{x}^{\prime} \cdot \acute{x}^{\prime}=0$ and $\acute{x}^{\prime\!\; 2}<0$  
are compatible with Eq. (\ref{2.25}) if and only if 
$\partial \sigma/\partial \tau^{\prime}=0$ and $\partial \sigma/\partial \sigma^{\prime}\neq0$. 
These imply that $\sigma$ is a strictly increasing or decreasing function only of $\sigma^{\prime}$, 
denoted as $\sigma=\sigma(\sigma^{\prime})$. 
With its inverse function, $\sigma^{\prime}=\sigma^{\prime}(\sigma)$, 
we have a restricted form of Eq. (\ref{2.4}): 
\begin{align}
\tau \rightarrow \tau^{\prime}=\tau^{\prime} (\tau, \sigma)\,, 
\qquad 
\sigma  \rightarrow \sigma^{\prime}=\sigma^{\prime} (\sigma)\,. 
\label{2.26}
\end{align}
As a result, it turns out that only the restricted reparametrization (\ref{2.26}) 
is allowed in the tensionless string, at least at the classical level, owing to the presence of the 
condition (\ref{2.1}). As we have seen,  
Eq. (\ref{2.26}) is obtained through the use of the classical equation (\ref{2.21}) derived from $S_{1}$, 
together with the additional condition (\ref{2.1}); 
in fact, Eq (\ref{2.26}) is found after solving Eqs. (\ref{2.1}) and (\ref{2.21}) simultaneously. 
For this reason, Eq. (\ref{2.26}) cannot be regarded as a primary transformation 
involved in the tensionless string model.  
The reparametrization that should first be considered at the action level is precisely   
the {\em full} reparametrization (\ref{2.4}); in fact, the action $S_{1}$ remains invariant  
under the reparametrization (\ref{2.4}).

From the Lagrangian (\ref{2.18}),  
the canonical momentum conjugate to $x^{\mu}$ is found to be 
\begin{align}
P_{\mu}:= \frac{\partial \mathcal{L}_{1}}{\partial \dot{x}^{\mu}}
=p_{\mu\nu} \acute{x}^{\nu}
=\frac{e}{e^{2}+f^{2}} \big\{ \acute{x}^{2} \dot{x}_{\mu} 
-(\dot{x}\cdot\acute{x}) \acute{x}_{\mu} \big\} \,, 
\label{2.27}
\end{align}
where Eq. (\ref{2.7}) has been used. Then, using Eq. (\ref{2.21}), we have 
\begin{align}
P_{\mu} P^{\mu}=0 \,.
\label{2.28}
\end{align}
(In the tensionful case, the use of Eq. (\ref{2.13}) leads to 
$P_{\mu} P^{\mu}+T^{2} \acute{x}^{2}=0$.) 
Also, the antisymmetry property of $p_{\mu\nu}$ in its indices guarantees that  
\begin{align}
\acute{x}^{\mu} P_{\mu}=0 \,.
\label{2.29}
\end{align}
Equations (\ref{2.28}) and (\ref{2.29}) are the Virasoro constraints characterizing 
the tensionless string.

\section{A spinor formulation of tensionless bosonic strings} 

The simultaneous equations (\ref{2.19}) and (\ref{2.20}) can be solved in terms of  
a space-time 2-component spinor  
$\bar{\pi}_{\alpha}=\bar{\pi}_{\alpha}(\tau, \sigma)$  $(\alpha=0, 1)$  
and its complex conjugate 
$\pi_{\dot{\alpha}}=\pi_{\dot{\alpha}}(\tau, \sigma)$ 
$(\dot{\alpha}=\dot{0}, \dot{1})$:   
\begin{align}
p_{\mu\nu}
&=\sigma_{\mu}{}^{\alpha\dot{\alpha}} \sigma_{\nu}{}^{\beta\dot{\beta}}  
p_{\alpha\beta\dot{\alpha}\dot{\beta}} 
\nonumber \\ 
&=\sigma_{\mu}{}^{\alpha\dot{\alpha}} \sigma_{\nu}{}^{\beta\dot{\beta}}  
\Big( \bar{\pi}_{\alpha} \bar{\pi}_{\beta} \epsilon_{\dot{\alpha} \dot{\beta}}  
+\pi_{\dot{\alpha}} \pi_{\dot{\beta}} \epsilon_{\alpha\beta} \Big)  \,. 
\label{3.1}
\end{align}
(For details, see Appendix A.) 
Here, $\bar{\pi}_{\alpha}$ and $\pi_{\dot{\alpha}}$ are assumed to behave 
as scalar fields on $\varXi$. 
The pair of Eqs. (\ref{2.19}) and (\ref{2.20}) is equivalent to Eq. (\ref{3.1}). 
As seen above, Eqs. (\ref{2.19}) and (\ref{2.20}) are incorporated in the 
Lagrangian (\ref{2.18}) by means of the auxiliary fields $e$ and $f$. 
To make a parallel formulation, we incorporate Eq. (\ref{3.1}) in a Lagrangian 
with the aid of Lagrange multiplier fields $h^{\mu\nu}=h^{\mu\nu}(\tau, \sigma)$ 
satisfying $h^{\mu\nu}=-h^{\nu\mu}$. 
They are assumed to transform in the same manner as $v^{\mu\nu}$ under 
the reparametrization (\ref{2.4}). An appropriate Lagrangian is given by 
\begin{align}
\mathcal{L}_{2}=\frac{1}{2} p_{\mu\nu} v^{\mu\nu}
-\frac{1}{2} h^{\mu\nu} \Big\{ p_{\mu\nu} 
-\sigma_{\mu}{}^{\alpha\dot{\alpha}} \sigma_{\nu}{}^{\beta\dot{\beta}}  
\Big( \bar{\pi}_{\alpha} \bar{\pi}_{\beta} \epsilon_{\dot{\alpha} \dot{\beta}}  
+\pi_{\dot{\alpha}} \pi_{\dot{\beta}} \epsilon_{\alpha\beta} \Big) \Big\} \,,
\label{3.2}
\end{align}
which can be expressed in the 2-component spinor notation as 
\begin{align}
\mathcal{L}_{2}=\frac{1}{2} p_{\alpha\beta\dot{\alpha}\dot{\beta}} 
v^{\alpha\beta\dot{\alpha}\dot{\beta}} 
-\frac{1}{2} h^{\alpha\beta\dot{\alpha}\dot{\beta}} 
\Big\{ p_{\alpha\beta\dot{\alpha}\dot{\beta}}  
-\Big(\bar{\pi}_{\alpha} \bar{\pi}_{\beta} \epsilon_{\dot{\alpha}\dot{\beta}}
+\pi_{\dot{\alpha}} \pi_{\dot{\beta}} \epsilon_{\alpha\beta} \Big)  \Big\} \,, 
\label{3.3}
\end{align}
where $v^{\alpha\beta\dot{\alpha}\dot{\beta}} 
=v^{\mu\nu}\sigma_{\mu}{}^{\alpha \dot{\alpha}} \sigma_{\nu}{}^{\beta \dot{\beta}}
=\dot{x}^{\alpha\dot{\alpha}} \acute{x}^{\beta\dot{\beta}} 
-\dot{x}^{\beta\dot{\beta}} \acute{x}^{\alpha\dot{\alpha}}$.    
From Eq. (\ref{3.3}), the Euler-Lagrange equation for 
$p_{\alpha\beta\dot{\alpha}\dot{\beta}}$ is found to be 
$h^{\alpha\beta\dot{\alpha}\dot{\beta}}=v^{\alpha\beta\dot{\alpha}\dot{\beta}}$. 
Putting it back to Eq. (\ref{3.3}), we have 
\begin{align}
\mathcal{L}_{2}& =\frac{1}{2} v^{\alpha\beta\dot{\alpha}\dot{\beta}}
\Big(\bar{\pi}_{\alpha} \bar{\pi}_{\beta} \epsilon_{\dot{\alpha}\dot{\beta}}
+\pi_{\dot{\alpha}} \pi_{\dot{\beta}} \epsilon_{\alpha\beta} \Big)
\nonumber 
\\ 
&=w^{\alpha\beta} \bar{\pi}_{\alpha} \bar{\pi}_{\beta}
+\bar{w}{}^{\dot{\alpha}\dot{\beta}} \pi_{\dot{\alpha}} \pi_{\dot{\beta}} 
\nonumber
\\
& =\dot{x}^{\alpha}{}_{\dot{\gamma}} \:\! 
\acute{x}^{\beta\dot{\gamma}} 
\bar{\pi}_{\alpha} \bar{\pi}_{\beta} 
+\dot{x}_{\gamma}{}^{\dot{\alpha}} \acute{x}^{\gamma\dot{\beta}}   
\pi_{\dot{\alpha}} \pi_{\dot{\beta}} \,, 
\label{3.4}
\end{align}
where 
\begin{align}
w^{\alpha\beta} :=
{1\over2} \Big(\dot{x}^{\alpha}{}_{\dot{\gamma}} \:\! \acute{x}^{\beta\dot{\gamma}} 
+\dot{x}^{\beta}{}_{\dot{\gamma}} \:\! \acute{x}^{\alpha\dot{\gamma}} \Big) \,.
\label{3.5}
\end{align}
Under the reparametrization (\ref{2.4}), $w^{\alpha\beta}$ transforms as 
\begin{align}
w^{\alpha\beta}
\rightarrow 
w^{\prime\!\:\alpha\beta}
=\left| \frac{\partial(\tau, \sigma)}{\partial(\tau^{\prime}, \sigma^{\prime})} \right| 
w^{\alpha\beta} \,. 
\label{3.6}
\end{align}
Hence, the action with the Lagrangian (\ref{3.4}), 
\begin{align}
S_{2}=\int_{\varXi} d\tau d\sigma \mathcal{L}_{2}  \,, 
\label{3.7}
\end{align}
remains reparametrization invariant.

Now, we demonstrate that the spinorial action $S_{2}$ actually describes the tensionless string. 
Variation of $S_{2}$ with respect to $\bar{\pi}_{\alpha}$ yields 
\begin{align}
w^{\alpha\beta} \bar{\pi}_{\beta}=0 \,. 
\label{3.8}
\end{align}
Taking into account the symmetry property $w^{\alpha\beta}=w^{\beta\alpha}$, 
we can solve Eq. (\ref{3.8}) as 
\begin{align}
w^{\alpha\beta}&=c\:\!\bar{\pi}^{\alpha} \bar{\pi}^{\beta} \,, 
\label{3.9}
\end{align}
where $c$ is a complex-valued scalar-density function on $\varXi$. 
We note here that $v^{\alpha\beta\dot{\alpha}\dot{\beta}}$ can be written as
\begin{align}
v^{\alpha\beta\dot{\alpha}\dot{\beta}}  
=w^{\alpha\beta} \epsilon^{\dot{\alpha}\dot{\beta}}
+\bar{w}^{\dot{\alpha}\dot{\beta}} \epsilon^{\alpha\beta} \,. 
\label{3.10}
\end{align}
(See Eq. (\ref{A11}) in Appendix A.) 
Then, using Eq. (\ref{3.9}) and  
$\bar{\pi}_{\alpha} \bar{\pi}{}^{\alpha}=\pi_{\dot{\alpha}} \pi^{\dot{\alpha}}=0$,  
it is readily seen that  
\begin{align}
\frac{1}{2} v_{\alpha\beta\dot{\alpha}\dot{\beta}}  
v^{\alpha\beta\dot{\alpha}\dot{\beta}} =
\big( c \:\! \bar{\pi}_{\alpha} \bar{\pi}{}^{\alpha} \big){}^{2} 
+\big( \bar{c} \:\! \pi_{\dot{\alpha}} \pi^{\dot{\alpha}} \big){}^{2} =0 \,. 
\label{3.11}
\end{align}
Equation (\ref{3.11}) turns out to be the null world-sheet condition (\ref{2.21}).  
For this reason, $S_{2}$ is recognized as an action for the tensionless string 
represented in a spinorial form.  
The action $S_2$ is essentially identical with the one discovered by 
Gusev and Zheltukhin in a simplistic way in a different context \cite{GZ}. 
In our approach, $S_2$ has been found in a systematical way based on the action (\ref{2.17}) 
by exactly solving the pair of constraint equations (\ref{2.19}) and (\ref{2.20}).

Substituting Eq. (\ref{3.9}) and its complex conjugate into Eq. (\ref{3.10}), we have  
\begin{align} 
v^{\alpha\beta\dot{\alpha}\dot{\beta}}  
=c\:\! \bar{\pi}^{\alpha} \bar{\pi}^{\beta} \epsilon^{\dot{\alpha}\dot{\beta}}
+\bar{c}\:\! \pi^{\dot{\alpha}} \pi^{\dot{\beta}} \epsilon^{\alpha\beta} \,. 
\label{3.12}
\end{align}
Here, we express the complex-valued function $c$ as $c=a+i b$, 
where $a$ and $b$ are real scalar-density functions on $\varXi$.  
Accordingly, Eq. (\ref{3.12}) can be written as 
\begin{align}
v^{\alpha\beta\dot{\alpha}\dot{\beta}} =a p^{\alpha\beta\dot{\alpha}\dot{\beta}} 
-b\!\: \tilde{p}^{\alpha\beta\dot{\alpha}\dot{\beta}} 
\label{3.13}
\end{align}
in terms of $p^{\alpha\beta\dot{\alpha}\dot{\beta}}
=\bar{\pi}^{\alpha} \bar{\pi}^{\beta} \epsilon^{\dot{\alpha}\dot{\beta}}
+\pi^{\dot{\alpha}} \pi^{\dot{\beta}} \epsilon^{\alpha\beta}$ 
and its Hodge dual $\tilde{p}^{\alpha\beta\dot{\alpha}\dot{\beta}}$.  
Setting $a=e$ and $b=-f$ in Eq. (\ref{3.13}), 
we see that Eq. (\ref{3.13}) just corresponds to Eq. (\ref{2.5}).  
In this way, Eq. (\ref{2.5}) can be reproduced from $S_{2}$.

Using the identity 
$2\acute{x}^{\alpha\dot{\beta}}  \acute{x}_{\alpha\dot{\gamma}}
=\acute{x}^{2} \delta^{\dot{\beta}}{}_{\dot{\gamma}}$, 
we can readily prove 
$2\acute{x}^{\alpha\dot{\alpha}} \acute{x}^{\beta\dot{\beta}} w_{\alpha\beta} 
=-\acute{x}^{2} \bar{w}^{\dot{\alpha}\dot{\beta}}$.  
This becomes 
\begin{align}
2c\:\! \acute{x}^{\alpha\dot{\alpha}} \bar{\pi}_{\alpha} 
\acute{x}^{\beta\dot{\beta}} \bar{\pi}_{\beta} 
=-\bar{c} \:\! \acute{x}^{2} \pi^{\dot{\alpha}} \pi^{\dot{\beta}}  
\label{3.14}
\end{align}
under the substitution of  Eq. (\ref{3.9}) and its complex conjugate. 
Contracting both sides of Eq. (\ref{3.14}) by $\pi_{\dot{\alpha}} \pi_{\dot{\beta}}$ 
leads to   
$2c\big(\acute{x}^{\alpha\dot{\alpha}} \bar{\pi}_{\alpha} \pi_{\dot{\alpha}}\big){}^{2}
=-\bar{c}\:\! \acute{x}^{2} \big(\pi_{\dot{\alpha}} \pi^{\dot{\alpha}}\big){}^{2}=0$.  
From this, provided $c\neq0$, it follows that 
\begin{align}
\acute{x}^{\alpha\dot{\alpha}} \bar{\pi}_{\alpha} \pi_{\dot{\alpha}}=0 \,, 
\label{3.15}
\end{align}
or equivalently 
\begin{align}
\acute{x}^{\alpha\dot{\alpha}} \bar{\pi}_{\alpha} =l \pi^{\dot{\alpha}} \,, 
\label{3.16}
\end{align}
where $l$ is a complex-valued function determined below. 
Combining Eq. (\ref{3.16}) and its complex conjugate, we obtain 
$\acute{x}^{\alpha\dot{\beta}}  \acute{x}_{\alpha\dot{\gamma}} \pi^{\dot{\gamma}}
=-|l|^{2} \pi^{\dot{\beta}}$. Using the identity 
$2\acute{x}^{\alpha\dot{\beta}}  \acute{x}_{\alpha\dot{\gamma}}
=\acute{x}^{2} \delta^{\dot{\beta}}{}_{\dot{\gamma}}$ again, 
$l$ is uniquely determined to be $l=\sqrt{-\acute{x}^{2}/2} \;\! e^{i\phi}$ 
up to a phase $\phi$. 
Here, we should note that this holds by virtue of the condition (\ref{2.1b}). 
Equation (\ref{3.16}) now reads 
\begin{align}
\acute{x}^{\alpha\dot{\alpha}} \bar{\pi}_{\alpha} =
\sqrt{-\frac{\acute{x}^{2}}{2}}\;\! e^{i\phi} 
\pi^{\dot{\alpha}} \,.
\label{3.17}
\end{align}
In this way, Eq. (\ref{3.17}) is derived from Eq. (\ref{3.15}). 
Conversely, Eq. (\ref{3.15}) is satisfied with Eq. (\ref{3.17}). 
For these reasons, the equivalence between Eqs. (\ref{3.15}) and (\ref{3.17}) is established. 
The phase $\phi$ seems to be arbitrary,  but actually, it is related to $e$ and $f$. 
In fact, substituting Eq. (\ref{3.17}) into Eq. (\ref{3.14}) gives $c=|c|e^{-i\phi}$, 
and hence, it follows that $\tan\phi=-b/a=f/e$. 
This implies that $\phi$ is a scalar field on $\varXi$ defined from 
the ratio of $f$ to $e$.  
The phase $\phi$ is thus fixed by means of Eq. (\ref{3.14}),  
which originates from the action $S_{2}$.

Recall that the null world-sheet condition (\ref{2.21}) was obtained also from 
the action $S_{2}$. 
As a result, Eqs. (\ref{2.22}) and (\ref{2.23}) are valid in the present spinor formulation. 
Following the procedure for deriving Eq. (\ref{3.17}), and using Eq. (\ref{2.22}),  
we can derive 
\begin{align}
\dot{x}^{\alpha\dot{\alpha}} \bar{\pi}_{\alpha} =0 
\label{3.18}
\end{align}
from the identity $2\dot{x}^{\alpha\dot{\beta}}  \dot{x}_{\alpha\dot{\gamma}}
=\dot{x}^{2} \delta^{\dot{\beta}}{}_{\dot{\gamma}}$ and Eq. (\ref{3.9}). 
Taking into account Eqs. (\ref{2.22}) and (\ref{2.23}),  
we can demonstrate that Eqs. (\ref{3.17}) and (\ref{3.18}) 
behave as a pair of covariant equations 
under the restricted reparametrization (\ref{2.26}), provided that 
$\sigma^{\prime}=\sigma^{\prime}(\sigma)$ is a strictly increasing function.

From the Lagrangian (\ref{3.4}), 
the canonical momentum conjugate to $x^{\alpha\dot{\alpha}}$ is found to be 
\begin{align}
P_{\alpha\dot{\alpha}}:= 
\frac{\partial \mathcal{L}_{2}}{\partial \dot{x}^{\alpha\dot{\alpha}}} 
=\acute{x}_{\beta\dot{\alpha}} \bar{\pi}_{\alpha} \bar{\pi}^{\beta} 
+\acute{x}_{\alpha\dot{\beta}} \pi_{\dot{\alpha}} \pi^{\dot{\beta}} \,, 
\label{3.19}
\end{align}
which yields  
\begin{align}
P^{2}
& :=P_{\alpha\dot{\alpha}} P^{\alpha\dot{\alpha}} 
=-2\big( \acute{x}^{\alpha\dot{\alpha}} \bar{\pi}_{\alpha} \pi_{\dot{\alpha}} \big){}^{2} \,,
\label{3.20}
\\
\acute{x}\cdot P
& :=\acute{x}^{\alpha\dot{\alpha}} P_{\alpha\dot{\alpha}} 
=\acute{x}^{\alpha\dot{\alpha}} \bar{\pi}_{\alpha} 
\acute{x}_{\beta\dot{\alpha}} \bar{\pi}^{\beta} 
+\acute{x}^{\alpha\dot{\alpha}} \pi_{\dot{\alpha}}  
\acute{x}_{\alpha\dot{\beta}} \pi^{\dot{\beta}} \,. 
\label{3.21}
\end{align}
As can be seen from Eq. (\ref{3.20}), the condition $P^{2}=0$ is equivalent to Eq. (\ref{3.15}),  
and, hence, to Eq. (\ref{3.17}).  
Also, substituting Eq. (\ref{3.17}) and its complex conjugate into Eq. (\ref{3.21}) leads to  
$\acute{x}\cdot P=0$. 
In this manner, $\acute{x}\cdot P=0$ can be obtained from $P^{2}=0$.  
This implies that in the present formulation, 
the Virasoro constraints $P^{2}=\acute{x}\cdot P=0$ eventually 
reduce to one of these constraints, $P^{2}=0$. 
Considering the equivalence between $P^{2}=0$ and Eq. (\ref{3.15}),  
we can treat Eq. (\ref{3.15}) or Eq. (\ref{3.17}) 
as an alternative form of the pair of the Virasoro constraints.

\section{An alternative spinorial action} 

In this section, we first derive a simpler equivalent form of the action $S_{2}$. 
Because the simpler form is no longer invariant under the full reparametrization (\ref{2.4}), 
next we improve it to an alternative spinorial action that remains invariant 
under the full reparametrization.

\subsection{A simpler equivalent form of the spinorial action $S_2$}

Applying Eq. (\ref{3.17}) and its complex conjugate to the Lagrangian (\ref{3.4}), we obtain 
\begin{align}
\tilde{\mathcal{L}}_{2}=-\sqrt{-2\acute{x}^2}\:\!\hat{e} \dot{x}^{\alpha\dot{\alpha}} 
\bar{\pi}_{\alpha} \pi_{\dot{\alpha}}\,, 
\label{4.1}
\end{align}
where $\hat{e}$ is a scalar field on $\varXi$ defined by 
$\hat{e}:=\cos\phi =e/\sqrt{e^{2}+f^{2}}$. 
Apart from the multiplicative factor $\sqrt{-2\acute{x}^2}\hat{e}$, 
Eq. (\ref{4.1}) has the same form as the spinorial Lagrangian 
for a massless spinless particle \cite{Shi}. 
This is compatible with the fact that each point on the tensionless string 
moves at the speed of light. 
Clearly, the action 
$\tilde{S}_{2}=\int_{\varXi} d\tau d\sigma \tilde{\mathcal{L}}_{2}$ 
is left invariant only under the very restricted class of reparametrizations, 
denoted by 
\begin{align}
\tau \rightarrow \tau^{\prime}=\tau^{\prime} (\tau)\,, 
\qquad 
\sigma  \rightarrow \sigma^{\prime}=\sigma^{\prime} (\sigma)\,. 
\label{4.2}
\end{align}
The invariance of the action $S_{2}$ under the full reparametrization (\ref{2.4}) is spoiled in  
the process of deriving Eq. (\ref{4.1}) from Eq. (\ref{3.4}).

It should be noted here that Eq. (\ref{4.1}) is not equivalent to Eq. (\ref{3.4}), 
because Eq. (\ref{4.1}) has been found through the use of Eq. (\ref{3.17}), and 
Eq. (\ref{3.17}) itself is not derivable from Eq. (\ref{4.1}). 
To extend Eq. (\ref{4.1}) so as to be equivalent to Eq. (\ref{3.4}), it is necessary 
to incorporate Eq. (\ref{3.17}) into Eq. (\ref{4.1}) at the Lagrangian level. 
For this purpose, noting the equivalence between Eqs. (\ref{3.15}) and (\ref{3.17}), 
we add $\acute{x}^{\alpha\dot{\alpha}} \bar{\pi}_{\alpha} \pi_{\dot{\alpha}}$ 
multiplied by a Lagrange multiplier field $\kappa=\kappa(\tau, \sigma)$ to Eq. (\ref{4.1}).  
The extension of Eq. (\ref{4.1}) is thus accomplished with 
\begin{align}
\mathcal{L}_{3}&=-\sqrt{-2\acute{x}^2}\:\! \hat{e} \dot{x}^{\alpha\dot{\alpha}} 
\bar{\pi}_{\alpha} \pi_{\dot{\alpha}}  
-\kappa \acute{x}^{\alpha\dot{\alpha}} \bar{\pi}_{\alpha} \pi_{\dot{\alpha}} 
\nonumber 
\\
&= -\Big( \sqrt{-2\acute{x}^2}\:\! \hat{e} \dot{x}^{\alpha\dot{\alpha}} 
+\kappa \acute{x}^{\alpha\dot{\alpha}} \Big) 
\bar{\pi}_{\alpha} \pi_{\dot{\alpha}} \,. 
\label{4.3}
\end{align}
Now, we verify the equivalence of the Lagrangians (\ref{3.4}) and (\ref{4.3}). 
Variation of the action 
\begin{align}
S_{3}=\int_{\varXi} d\tau d\sigma \mathcal{L}_{3}  
\label{4.4}
\end{align}
with respect to $\hat{e}$ and $\kappa$ yields 
\begin{align}
\dot{x}^{\alpha\dot{\alpha}} \bar{\pi}_{\alpha} \pi_{\dot{\alpha}}=0 
\label{4.5} 
\end{align}
and Eq. (\ref{3.15}). Here, the condition $\acute{x}^2 <0$ has been taken into account.  
Equation (\ref{3.15}), which is now understood as a Euler-Lagrange equation derived 
from $S_{3}$, leads to Eq. (\ref{3.17}) with a phase that, in general, is different from 
$\phi$ such that $\tan\phi=f/e$. 
(Recall that $\phi$ was fixed by using an equation obtained from the action (\ref{3.7}).) 
To state this situation clearly, we rewrite Eq. (\ref{3.17}) as 
$\acute{x}^{\alpha\dot{\alpha}} \bar{\pi}_{\alpha} =
\sqrt{-\acute{x}^{2}/2}\;\! e^{i\varphi} \pi^{\dot{\alpha}}$, 
with an arbitrary real function $\varphi$ on $\varXi$, instead of $\phi$.  
Applying this expression and its complex conjugate to Eq. (\ref{4.3}),  
we can immediately show that Eq. (\ref{4.3}) becomes Eq. (\ref{3.4}) 
after carrying out the phase transformation 
$\bar{\pi}_{\alpha} \rightarrow  e^{-i(\phi-\varphi)/2} \bar{\pi}_{\alpha}$.  
Conversely, as shown above, Eq. (\ref{4.3}) can be found from Eq. (\ref{3.4}). 
For these reasons, 
the classical equivalence of the Lagrangians (\ref{3.4}) and (\ref{4.3}) is established;  
$S_{3}$ is now regarded as a simpler equivalent form of $S_{2}$.  
In spite of this classical equivalence,  
the symmetries of $S_{2}$ and $S_{3}$ are different: 
The action $S_{2}$ is invariant under the full reparametrization (\ref{2.4}), 
while the action $S_{3}$ is invariant only under the restricted reparametrization 
(\ref{4.2}), provided that $\kappa$ transforms as 
$\kappa\rightarrow \kappa^{\prime}
=(\partial\tau/\partial\tau^{\prime}) \kappa$. 
Such a difference in symmetries between equivalent actions is encountered  
also in some of the gauge theories.

Variation of the action $S_{3}$ with respect to $\pi_{\dot{\alpha}}$ gives 
\begin{align}
\Big( \sqrt{-2\acute{x}^2}\:\! \hat{e} \dot{x}^{\alpha\dot{\alpha}} 
+\kappa \acute{x}^{\alpha\dot{\alpha}} \Big) \bar{\pi}_{\alpha}=0\,. 
\label{4.6}
\end{align}
Because 
$\sqrt{-2\acute{x}^2}\:\!\hat{e} \dot{x}^{\alpha\dot{\alpha}} 
+\kappa \acute{x}^{\alpha\dot{\alpha}}$ is Hermitian, 
Eq. (\ref{4.6}) can be solved as 
\begin{align}
\sqrt{-2\acute{x}^2}\:\!\hat{e} \dot{x}^{\alpha\dot{\alpha}} 
+\kappa \acute{x}^{\alpha\dot{\alpha}}=u \bar{\pi}^{\alpha} \pi^{\dot{\alpha}} \,,
\label{4.7}
\end{align}
with $u$ being a real function on $\varXi$. 
(Recall here that $x^{\alpha\dot{\alpha}}$ is Hermitian, i.e.,   
$\overline{x^{\beta\dot{\alpha}}}=x^{\alpha\dot{\beta}}$,  
because $x^{\mu}$ is real.)  
Contracting both sides of Eq. (\ref{4.7}) by $\dot{x}_{\alpha\dot{\alpha}}$ 
and using Eq. (\ref{4.5}), we have 
\begin{align}
\sqrt{-2\acute{x}^2}\:\!\hat{e} \dot{x}{}^{2}+\kappa \dot{x}\cdot\acute{x}=0 \,.
\label{4.8}
\end{align}
Similarly, contracting both sides of Eq. (\ref{4.7}) by $\acute{x}_{\alpha\dot{\alpha}}$ 
and using Eq. (\ref{3.15}) yield
\begin{align}
\sqrt{-2\acute{x}^2}\:\!\hat{e} \dot{x}\cdot\acute{x}+\kappa \acute{x}{}^{2}=0 \,.
\label{4.9}
\end{align}
Equations (\ref{4.8}) and (\ref{4.9}), which are treated as a set of 
simultaneous equations in $\hat{e}$ and $\kappa$, can possess nontrivial solutions, 
if and only if $\dot{x}^{2} \acute{x}^{2} -(\dot{x} \cdot \acute{x})^{2}=0$. 
This relation is precisely the null world-sheet condition (\ref{2.21}). 
Since $S_{3}$ gives Eq. (\ref{2.21}) in this way, 
it is recognized as an action for the tensionless string. 
As already shown, the condition (\ref{2.21}), together with the condition (\ref{2.1}), 
leads to Eqs. (\ref{2.22}) and (\ref{2.23}).  
Substituting these into Eqs. (\ref{4.8}) and (\ref{4.9}), we have $\kappa=0$,  
and accordingly, Eq. (\ref{4.3}) reduces to Eq. (\ref{4.1}). 
Here, it should be stressed that this reduction to Eq. (\ref{4.1}) is seen 
as the result of solving the Euler-Lagrange equations derived from $S_{3}$.

From the Lagrangian (\ref{4.3}), 
the canonical momentum conjugate to $x^{\alpha\dot{\alpha}}$ is found to be   
$P_{\alpha\dot{\alpha}}=-\sqrt{-2\acute{x}^2} \hat{e} 
\bar{\pi}_{\alpha} \pi_{\dot{\alpha}}$. 
Then, $P^{2}=0$ is automatically satisfied, while  
the condition $\acute{x}\cdot P=0$ is valid with Eq. (\ref{3.15}).  
In this manner, the Virasoro constraints for the tensionless string are verified also 
in the formulation based on $S_{3}$.

\subsection{A fully reparametrization-invariant spinorial action}

Let us introduce a real contravariant vector-density field 
$\lambda^{j}=\lambda^{j}(\tau, \sigma)$ $(j=0,1)$ on $\varXi$,    
which transforms under the reparametrization (\ref{2.4}) as 
\begin{align}
\lambda^{j}(\xi) \rightarrow 
\lambda^{\prime\!\: j} (\xi^{\prime})
=\left| \frac{\partial \xi}{\partial \xi^{\prime}} \right| 
\frac{\partial \xi^{\prime\!\: j}}{\partial \xi^{k}} \lambda^{k} (\xi)\,, 
\label{4.10}
\end{align}
where $(\xi^{0}, \xi^{1}):=(\tau, \sigma)$. 
Replacing $\sqrt{-2\acute{x}^2} \hat{e}$ and $\kappa$ in Eq. (\ref{4.4}) by 
$\lambda^{0}$ and $\lambda^{1}$, respectively, 
we define the action  
\begin{align}
S_{4}=\int_{\varXi} d^2 \xi \mathcal{L}_{4}  
\label{4.11}
\end{align}
with the Lagrangian 
\begin{align}
\mathcal{L}_{4}&=-\lambda^{0} \dot{x}^{\alpha\dot{\alpha}} 
\bar{\pi}_{\alpha} \pi_{\dot{\alpha}} 
-\lambda^{1}
\acute{x}^{\alpha\dot{\alpha}} \bar{\pi}_{\alpha} \pi_{\dot{\alpha}} 
\nonumber 
\\
&=-\lambda^{j} \big( \partial_{j} x^{\alpha\dot{\alpha}} \big) 
\bar{\pi}_{\alpha} \pi_{\dot{\alpha}} \,,
\label{4.12}
\end{align}
where $\partial_{j}:=\partial/\partial \xi^{j}$. 
It is evident that unlike the action $S_{3}$, the action $S_{4}$ remains invariant under 
the full reparametrization (\ref{2.4}). In this sense, $S_{4}$ is an improved version of $S_{3}$; 
the action $S_{3}$ may be regarded as $S_{4}$ in a particular gauge 
$(\lambda^{0}, \lambda^{1})=\big( \sqrt{-2\acute{x}^2} \hat{e}, \kappa \big)$. 
Varying $\lambda^{j}$ in $S_{4}$, 
we have $\big(\partial_{j} x^{\alpha\dot{\alpha}} \big)\bar{\pi}_{\alpha} \pi_{\dot{\alpha}}=0$ 
or, equivalently, Eqs. (\ref{4.5}) and (\ref{3.15}). 
Also, the equation given by varying $\pi_{\dot{\alpha}}$ in $S_{4}$ can be 
solved as 
$\lambda^{j} \partial_{j} x^{\alpha\dot{\alpha}}=u \bar{\pi}^{\alpha} \pi^{\dot{\alpha}}$. 
Combining this and 
$\big(\partial_{j} x^{\alpha\dot{\alpha}} \big)\bar{\pi}_{\alpha} \pi_{\dot{\alpha}}=0$, 
we obtain the simultaneous equations 
$\lambda^{j} \big(\partial_{j} x^{\alpha\dot{\alpha}} \big) \partial_{k} x_{\alpha\dot{\alpha}}=0$ 
$(k=0,1)$ in $\lambda^{j}$.  
These are analogs of Eqs. (\ref{4.8}) and (\ref{4.9}), and 
lead to the null world-sheet condition (\ref{2.21}), provided that 
the trivial solution $\lambda^{0}=\lambda^{1}=0$ is not admitted. 
In this way, we can derive Eq. (\ref{2.21}) from the action $S_{4}$, and therefore,   
$S_{4}$ is confirmed to be an alternative spinorial action for the tensionless string.

The spinorial action $S_{2}$ can be found directly from $S_{4}$ in the following manner: 
Variation of $S_{4}$ with respect to $\lambda^{1}$ yields Eq. (\ref{3.15}), 
which, as shown above, leads to 
$\acute{x}^{\alpha\dot{\alpha}} \bar{\pi}_{\alpha} =
\sqrt{-\acute{x}^{2}/2}\;\! e^{i\varphi} \pi^{\dot{\alpha}}$.   
Applying this to Eq. (\ref{4.12}), we have 
\begin{align}
\mathcal{L}_{4}
=\frac{\lambda^{0}}{\sqrt{-2\acute{x}^{2}}} 
\Big( e^{-i\varphi} 
\dot{x}^{\alpha}{}_{\dot{\gamma}} \:\! 
\acute{x}^{\beta\dot{\gamma}} 
\bar{\pi}_{\alpha} \bar{\pi}_{\beta} 
+e^{i\varphi} 
\dot{x}_{\gamma}{}^{\dot{\alpha}} \acute{x}^{\gamma\dot{\beta}}   
\pi_{\dot{\alpha}} \pi_{\dot{\beta}} \Big) \,.
\label{4.12.1}
\end{align}
Here, we impose the condition $\lambda^{0}=\sqrt{-2\acute{x}^{2}}$ on Eq. (\ref{4.12.1}) 
so that the invariance of $S_{4}$ under the full reparametrization (\ref{2.4}) can be restored. 
Then, by performing the phase transformation 
$\bar{\pi}_{\alpha} \rightarrow e^{i\varphi/2} \bar{\pi}_{\alpha}$,  
Eq. (\ref{4.12.1}) reduces to Eq. (3.4). Thus, we obtain the action $S_{2}$.

Now, we define a twistor $X^{A}$ and its dual twistor $\bar{X}_{A}$ by   
\begin{align}
X^{A}=\big(\chi^{\alpha}, \pi_{\dot{\alpha}} \big)\,,  
\qquad 
\bar{X}_{A}=\big(\bar{\pi}_{\alpha}, \bar{\chi}^{\dot{\alpha}} \big) \,, 
\label{4.13}
\end{align}
where $\chi^{\alpha}$ and $\bar{\chi}^{\dot{\alpha}}$ 
are space-time 2-component spinors defined by    
\begin{align}
\chi^{\alpha}=ix^{\alpha\dot{\alpha}} \pi_{\dot{\alpha}} \,, 
\qquad 
\bar{\chi}^{\dot{\alpha}}=-ix^{\alpha\dot{\alpha}} \bar{\pi}_{\alpha} \,.
\label{4.14}
\end{align}
With Eq. (\ref{4.14}), it is readily shown that $X^{A}$ is a null twistor: 
\begin{align}
\bar{X}_{A}X^{A}
=\bar{\pi}_{\alpha} \chi^{\alpha}+\bar{\chi}^{\dot{\alpha}}\pi_{\dot{\alpha}}
=0 \,. 
\label{4.15}
\end{align}
The twistor components $X^{A}$ and $\bar{X}_{A}$ $(A=0,1,2,3)$ 
are treated as scalar fields on $\varXi$, 
since $\pi_{\dot{\alpha}}$, $\bar{\pi}_{\alpha}$ and 
$x^{\alpha\dot{\alpha}}$ have been assumed to be scalar fields on $\varXi$. 
In terms of $X^{A}$ and $\bar{X}_{A}$, the Lagrangian (\ref{4.12}) is written as
\begin{align}
\mathcal{L}_{4}=\frac{i}{2} \lambda^{j} 
\big(\bar{X}_{A}\partial_{j} X^{A} -X^{A}\partial_{j} \bar{X}_{A} \big) \,.
\label{4.16}
\end{align}

\section{A twistor formulation of tensionless bosonic strings}

In the picture of twistor theory, 
space-time points are taken to be a secondary construct, and twistors 
(or twistor components) are regarded as more primitive variables 
than the space-time coordinates. 
Following this, first, we introduce twistors 
\begin{align}
Z^{A}=\big( \omega^{\alpha}, \pi_{\dot{\alpha}} \big)\,,  
\qquad 
\bar{Z}_{A}=\big( \bar{\pi}_{\alpha}, \bar{\omega}^{\dot{\alpha}} \big) \,,
\label{5.1}
\end{align}
without referring to Eq. (\ref{4.14}). 
For a short while,  
it is assumed that $Z^{A}$ and $\bar{Z}_{A}$ are free of the null twistor condition 
\begin{align}
\bar{Z}_{A} Z^{A}=0 \,, 
\label{5.2}
\end{align}
and accordingly, the components $Z^{A}$ $(A=0,1,2,3)$ can take any complex values  
independently. Simply replacing $X^{A}$ and $\bar{X}_{A}$ in Eq. (\ref{4.16}) by $Z^{A}$  
and $\bar{Z}_{A}$, respectively, we define the Lagrangian
\begin{align}
\mathcal{L}_{5}=\frac{i}{2} \lambda^{j} 
\big( \bar{Z}_{A}\partial_{j} Z^{A} -Z^{A}\partial_{j} \bar{Z}_{A} \big) \,. 
\label{5.3}
\end{align}
Here, the components $Z^{A}$ and $\bar{Z}_{A}$ are understood as scalar fields 
on $\varXi$, so that the action 
\begin{align}
S_5=\int_{\Xi} d^2 \xi \mathcal{L}_5 
\label{5.3.1}
\end{align}
is reparametrization invariant.   
Mathematically, the scalar fields $Z^{A}=Z^{A}(\xi)$ define 
a map from the parameter space $\varXi$ to the twistor space $\mathbf{T}$,  
a complex 4-dimensional vector space coordinatized by 
$\big(Z^{A}\big):=\big(Z^{0}, Z^{1}, Z^{2}, Z^{3} \big)$.  
As is well known in twistor theory \cite{PM, Pen2, PR, Pen, HT, Tak},  
a point in $\mathbf{T}$, denoted by $\big(Z^{A}\big)$ with $\pi_{\dot{\alpha}}\neq 0$, 
corresponds to an $\alpha$-plane, a totally null complex 2-plane,  
in complexified Minkowski space $\Bbb{C}\mathbf{M}$.\footnote{
The restriction $\pi_{\dot{\alpha}}\neq 0$ can be removed   
if we take into account the $\alpha$-planes in complexified {\em compactified}  
Minkowski space $\Bbb{C}\mathbf{M}^{\sharp}$, 
not only those in $\Bbb{C}\mathbf{M}$. In the present paper, however, 
we assume $\pi_{\dot{\alpha}}\neq 0$, and accordingly, do not consider 
$\Bbb{C}\mathbf{M}^{\sharp}$.} 
More precisely, 
for an arbitrary twistor $Z^{A}=(\omega^{\alpha}, \pi_{\dot{\alpha}})$ 
with $\pi_{\dot{\alpha}}\neq0$, there exists an $\alpha$-plane
\begin{align}
\alpha_{Z}:=\big\{ \big(z^{\alpha\dot{\alpha}}\big) \in \Bbb{C}\mathbf{M}\, | 
\:\omega^{\alpha}=iz^{\alpha\dot{\alpha}} \pi_{\dot{\alpha}} \big\}\,.
\label{5.4}
\end{align}

Suppose now that the twistor $Z^{A}$ satisfies the condition (\ref{5.2}).  
Then, the equation 
\begin{align}
\omega^{\alpha}=iz^{\alpha\dot{\alpha}} \pi_{\dot{\alpha}} 
\label{5.5}
\end{align}
has an Hermitian solution $z^{\alpha\dot{\alpha}}=x^{\alpha\dot{\alpha}}$.\footnote{
The Hermitian solution is actually given by \cite{PR, HT}
$$x^{\alpha\dot{\alpha}}=-i \Big( \bar{\omega}{}^{\dot{\beta}} \pi_{\dot{\beta}} \Big)^{-1} 
\omega^{\alpha} \bar{\omega}^{\dot{\alpha}} +\ell \:\!\bar{\pi}^{\alpha} \pi^{\dot{\alpha}}\,, 
\quad \ell \in \Bbb{R}\,.$$
It is worthwhile mentioning here that 
$Z^{A}$ with $\bar{\omega}{}^{\dot{\beta}} \pi_{\dot{\beta}}\neq0$ 
can always be provided 
by utilizing the degrees of freedom of choosing the origin in $\mathbf{M}$. 
}  
In other words, if $Z^{A}$ is a null twistor, 
the corresponding $\alpha$-plane, $\alpha_{Z}$, contains real points, which constitute 
a line of intersection with Minkowski space $\mathbf{M}$. 
This line is precisely a null geodesic in $\mathbf{M}$. 
(If $\bar{Z}_{A}Z^{A}\neq0$, Eq. (\ref{5.5}) has no Hermitian solutions, and hence,  
$\alpha_{Z}$ contains no real points.)  
The general solution of Eq. (\ref{5.5}) that includes the solution  
$z^{\alpha\dot{\alpha}}=x^{\alpha\dot{\alpha}}$ is given by   
\begin{align}
z^{\alpha\dot{\alpha}}=x^{\alpha\dot{\alpha}}+\bar{\iota}^{\,\alpha} \pi^{\dot{\alpha}}, 
\label{5.6}
\end{align}
where $\bar{\iota}^{\,\alpha}$ is an arbitrary spinor.  
Obviously, Eq. (\ref{5.6}) defines an $\alpha$-plane that intersects with $\mathbf{M}$. 
Because the twistor components $Z^{A}$ are now scalar fields on $\Xi$,  
all the constituents of Eq. (\ref{5.6}) are treated as scalar fields on $\Xi$. 
Using Eqs. (\ref{5.5}) and (\ref{5.6}), we can prove 
\begin{align}
&\bar{Z}_{A}\partial_{j} Z^{A} -Z^{A}\partial_{j} \bar{Z}_{A} 
\nonumber \\
&=i \bar{\pi}_{\alpha} \pi_{\dot{\alpha}}
\partial_{j} \big(z^{\alpha\dot{\alpha}}+\bar{z}^{\alpha\dot{\alpha}}\big) 
+i \big( z^{\alpha\dot{\alpha}}-\bar{z}^{\alpha\dot{\alpha}} \big) 
( \bar{\pi}_{\alpha} \partial_{j} \pi_{\dot{\alpha}}
-\pi_{\dot{\alpha}} \partial_{j} \bar{\pi}_{\alpha} )
\nonumber \\
&=2i \big( \partial_{j} x^{\alpha\dot{\alpha}} \big) \bar{\pi}_{\alpha} \pi_{\dot{\alpha}} \,, 
\label{5.7}
\end{align}
which demonstrates that the Lagrangian $\mathcal{L}_{5}$,  
subjected to the condition (\ref{5.2}),  reduces to the Lagrangian (\ref{4.12}). 
Then, we see that to describe the tensionless string in terms of $Z^{A}$ and $\bar{Z}_{A}$ 
at the Lagrangian level, it is necessary to incorporate Eq. (\ref{5.2}) into $\mathcal{L}_{5}$. 
The incorporation is accomplished by adding $\bar{Z}_{A} Z^{A}$ multiplied  
by a Lagrange multiplier field $\rho=\rho(\xi)$ to $\mathcal{L}_{5}$.  
The desirable Lagrangian is thus given by  
\begin{align}
\mathcal{L}_{6}=\frac{i}{2} \lambda^{j} 
\big(\bar{Z}_{A}\partial_{j} Z^{A} -Z^{A}\partial_{j} \bar{Z}_{A} \big) 
+\rho \bar{Z}_{A} Z^{A}\,. 
\label{5.8}
\end{align}
Here, $\rho$ is assumed to be a real scalar-density field on $\Xi$ so that the action 
\begin{align}
S_6=\int_{\Xi} d^2 \xi \mathcal{L}_6 
\label{5.9}
\end{align}
can be reparametrization invariant. Variation of $S_{6}$ with respect to $\rho$ yields 
the condition (\ref{5.2}), and consequently, $S_{6}$ turns out to be equivalent to 
the spinorial action $S_{4}$. 
The action $S_{6}$ is therefore recognized as a twistorial action for the tensionless string.

\section{Local internal symmetries in the twistor formulation}

In the previous section, the condition (\ref{5.2}) has been incorporated into the Lagrangian 
$\mathcal{L}_{5}$ in an ad hoc manner with the aid of the Lagrange multiplier field $\rho$. 
Here, we show that Eq. (\ref{5.2}) and the Lagrangian (\ref{5.8}) can be found 
automatically and naturally on the basis of a phase symmetry inherent in $\mathcal{L}_{5}$.

We first note that $\mathcal{L}_{5}$ remains invariant under the phase transformation 
\begin{align}
Z^{A} \rightarrow Z^{\prime A}=e^{i\theta} Z^{A} \,, \qquad 
\bar{Z}_{A} \rightarrow \bar{Z}^{\prime}_{A}=e^{-i\theta} \bar{Z}_{A} \,. 
\label{6.1}
\end{align}
At this stage, $\theta$ is assumed to be a real constant. 
Now, we apply the so-called gauge principle to the present formulation. 
In accordance with this principle, the phase transformation (\ref{6.1}) 
must be carried out at each point on the parameter space $\varXi$ independently, 
while preserving the smoothness of the transformed fields  
$Z^{\prime A}=Z^{\prime A}(\xi)$. 
This is realized by replacing the constant $\theta$ with a real smooth function 
$\theta(\xi)$ on $\varXi$. After the replacement, Eq. (\ref{6.1}) is read as a 
{\em local} phase transformation, 
and no longer leaves $\mathcal{L}_{5}$ invariant. 
To find an invariant Lagrangian, 
following the usual procedure in gauge theories, we introduce a $U(1)$ gauge field,  
$a_{j}=a_{j}(\xi)$, and the associated covariant derivative 
$D_{j}:=\partial_{j}-i a_{j}$ into $\varXi$. 
Then, replacing $\partial_{j}$ in Eq. (\ref{5.3}) by $D_{j}$, we define 
\begin{align}
\mathcal{L}_{7}=\frac{i}{2} \lambda^{j} 
\big(\bar{Z}_{A} D_{j} Z^{A} -Z^{A} \bar{D}_{j} \bar{Z}_{A} \big) \,. 
\label{6.2}
\end{align}
This Lagrangian remains invariant under the transformation (\ref{6.1}) 
supplemented by the gauge transformation 
\begin{align}
a_{j} \rightarrow a_{j}^{\prime}=a_{j}+\partial_{j} \theta \,.
\label{6.3}
\end{align}
It is remarkable that $\mathcal{L}_{7}$ reduces to the Lagrangian (\ref{5.8}) 
by setting $\rho=\lambda^{j} a_{j}$. 
The gauge transformation of $\rho$ is determined to be  
\begin{align}
\rho \rightarrow \rho^{\prime}=\rho+d_{\lambda} \theta \,,
\label{6.4}
\end{align}
where $d_{\lambda}$ denotes the directional derivative 
$d_{\lambda}:=\lambda^{j} \partial_{j}$. 
Obviously, the Lagrangian (\ref{5.8}) is left invariant under 
the simultaneous transformations (\ref{6.1}) and (\ref{6.4}). 
Varying $a_{j}$ in the action  
\begin{align}
S_7=\int_{\Xi} d^2 \xi \mathcal{L}_7 \,, 
\label{6.4.1}
\end{align}
we have $\lambda^{j} \bar{Z}_{A} Z^{A}=0$,  
which leads to Eq. (\ref{5.2}) under the condition 
$( \lambda^{0}, \lambda^{1} ) \neq (0, 0)$ 
necessary for avoiding trivialities. 
Thus, Eq. (\ref{5.2}) and the Lagrangian (\ref{5.8}) 
can be found from $\mathcal{L}_{5}$ on the basis of the gauge principle.

As readily seen, the Lagrangian (\ref{5.8}) remains invariant under the {\em local} scale 
transformation 
\begin{subequations}
\label{6.5}
\begin{alignat}{3}
Z^{A} &\rightarrow Z^{\prime A}=r Z^{A} \,, 
&\qquad 
\bar{Z}_{A} &\rightarrow \bar{Z}^{\prime}_{A}=r \bar{Z}_{A} \,, 
\label{6.5a}
\\
\lambda^{j} &\rightarrow \lambda^{\prime\!\: j}=r^{-2} \lambda^{j} \,, 
&\qquad
\rho &\rightarrow \rho^{\prime}=r^{-2} \rho \,, 
\label{6.5b}
\end{alignat}
\end{subequations}
where $r$ is a positive-valued smooth function on $\varXi$. 
The transformations (\ref{6.1}) and (\ref{6.5a}) can be combined as  
a {\em complexified} local scale transformation 
\begin{align}
Z^{A} &\rightarrow Z^{\prime A}=\upsilon Z^{A} \,, 
\qquad 
\bar{Z}_{A} \rightarrow \bar{Z}^{\prime}_{A}
=\bar{\upsilon} \bar{Z}_{A} \,, 
\label{6.6}
\end{align}
where $\upsilon:=re^{i\theta}$. 
Now, we can say that the Lagrangian (\ref{5.8}) remains invariant  
under the transformation (\ref{6.6}) supplemented by the transformations 
\begin{align}
\lambda^{j} \rightarrow \lambda^{\prime\!\: j}=r^{-2} \lambda^{j} \,, 
\qquad
\rho \rightarrow 
\rho^{\prime}=r^{-2} (\rho+d_{\lambda} \theta) \,.
\label{6.7}
\end{align}
This implies that actually 
the Lagrangian (\ref{5.8}) is defined for the proportionality class,    
referred to as the {\em projective} twistor,  
$[ Z^{A} ] :=
\big\{ \upsilon Z^{A} \big|\, \upsilon \in \Bbb{C}\setminus\{0\} \big\}$ 
rather than the (nonzero) twistor $Z^{A}$ itself.  
(Recall here that $\pi_{\dot{\alpha}}\neq 0$ has been assumed.)  
Correspondingly, $Z^{A}$ in Eq. (\ref{5.8}) should be understood as scalar fields  
that define a map from $\varXi$ to the projective twistor space \cite{PR, Pen, HT, Tak} 
\begin{align}
\mathbf{PT}:= 
\big\{ \big[ \big( Z^{A} \big) \big]\, \big|\, \big( Z^{A} \big) 
\in \mathbf{T} \setminus\{(0,0,0,0)\} \big\}\,, 
\label{6.8}
\end{align}
where 
$\big[ \big( Z^{A} \big) \big] :=
\big\{ \big( \upsilon Z^{A} \big) \big|\, 
\upsilon\in \Bbb{C}\setminus\{0\} \big\}$.   
Our twistor formulation is thus described with the projective twistor $[ Z^{A} ]$. 
This result is consistent with the fact that in twistor theory,  
$\mathbf{PT}$ is treated as a space more essential than $\mathbf{T}$. 
The projective twistor $[ Z^{A} ]$ is also expressed as $Z^{A}$ by regarding  
it as a representative element of the set $[ Z^{A} ]$.

\section{Some classical analyses in the twistor formulation}

We can assume, without loss of generality, that $\lambda^{0} >0$, 
provided the trivial situation $\lambda^{0}=0$ is excluded. 
Then, in terms of the (rescaled) twistors 
$\mathscr{Z}^{A}:=\sqrt{\lambda^{0}} Z^{A}$ and 
$\bar{\mathscr{Z}}_{A}:=\sqrt{\lambda^{0}} \bar{Z}_{A}$  
$\,(\lambda^{0} >0)$, 
the Lagrangian (\ref{5.8}) can be written as 
\begin{align}
\mathcal{L}_{6}=
\frac{i}{2} \Big(\bar{\mathscr{Z}}_{A} \dot{\mathscr{Z}}^{A} 
-\mathscr{Z}^{A} \dot{\bar{\mathscr{Z}}}_{A} \Big) 
+\frac{i}{2} \hat{\lambda}
\Big(\bar{\mathscr{Z}}_{A} \acute{\mathscr{Z}}^{A} 
-\mathscr{Z}^{A} \acute{\bar{\mathscr{Z}}}_{A} \Big)
+\varrho \!\:\bar{\mathscr{Z}}_{A} \mathscr{Z}^{A} , 
\label{7.1}
\end{align}
where $\hat{\lambda}:=\lambda^{1}/\lambda^{0}$ and 
$\varrho:=\rho/\lambda^{0}$. 
The twistor $\mathscr{Z}^{A}$ is precisely a representative element of  
$[ Z^{A} ]$. Under the full reparametrization (\ref{2.4}), the fields 
$\mathscr{Z}^{A}$, $\hat{\lambda}$ and $\varrho$ transform according to 
the complicated rules that are 
defined from the transformation rules of $Z^{A}$, $\lambda^{j}$ and $\rho$. 
By virtue of this, 
the action $S_6=\int_{\Xi} d^2 \xi \mathcal{L}_6$ with the Lagrangian (\ref{7.1}) 
remains invariant under the full reparametrization.   
The gauge transformation (\ref{6.4}) reads 
\begin{align}
\varrho \rightarrow \varrho^{\prime}=
\varrho+\dot{\theta}+\hat{\lambda} \acute{\theta} \,.
\label{7.2}
\end{align}
Variation of the action $S_6$ with respect to 
$\bar{\mathscr{Z}}_{A}$ gives the equation of motion 
\begin{align}
\dot{\mathscr{Z}}^{A}+\hat{\lambda} \acute{\mathscr{Z}}^{A}
+\bigg(\!-i\varrho +\frac{1}{2} \Acute{\Hat{\lambda}} \bigg)
\mathscr{Z}^{A}=0 \,.
\label{7.3}
\end{align}
In addition, varying $\hat{\lambda}$ and $\varrho$ in $S_6$ yields     
\begin{align}
\mathcal{R} &:=\frac{i}{2} 
\Big(\bar{\mathscr{Z}}_{A} \acute{\mathscr{Z}}^{A} 
-\mathscr{Z}^{A} \acute{\bar{\mathscr{Z}}}_{A} \Big)=0 \,, 
\label{7.4}
\\
\mathcal{S} &:=\frac{1}{2} \bar{\mathscr{Z}}_{A} \mathscr{Z}^{A} =0\,.
\label{7.5}
\end{align}
Using Eqs. (\ref{7.3}) and (\ref{7.4}), we can show that 
\begin{align}
\bar{\mathscr{Z}}_{A} \dot{\mathscr{Z}}^{A} 
-\mathscr{Z}^{A} \dot{\bar{\mathscr{Z}}}_{A}=0 \,.
\label{7.6}
\end{align}
Because $\mathscr{Z}^{A}$ satisfies the null twistor condition (\ref{7.5}),  
there exist Hermitian coordinate variables 
$x^{\alpha\dot{\beta}}\!\: \big(=\overline{x^{\beta\dot{\alpha}}} \,\big)$  
such that 
$\mathscr{Z}^{A}=\big( ix^{\alpha\dot{\beta}} \varpi_{\dot{\beta}}, \varpi_{\dot{\alpha}} \big)$. 
In terms of $x^{\alpha\dot{\alpha}}$ and $\varpi_{\dot{\alpha}}$, the conditions 
(\ref{7.4}) and (\ref{7.6}) are written together as  
$\big( \partial_{j} x^{\alpha\dot{\alpha}} \big) \bar{\varpi}_{\alpha} \varpi_{\dot{\alpha}}=0$. 
Equation (\ref{7.3}) leads to 
$\dot{x}^{\alpha\dot{\alpha}} +\hat{\lambda} \acute{x}^{\alpha\dot{\alpha}} 
=u \bar{\varpi}^{\alpha} \varpi^{\dot{\alpha}}$.   
Combining these equations, we have the pair of equations  
$\big( \dot{x}^{\alpha\dot{\alpha}} +\hat{\lambda} \acute{x}^{\alpha\dot{\alpha}} \big)
\partial_{j} x_{\alpha\dot{\alpha}}=0$ $(j=0, 1)$, which yields 
$\dot{x}{}^{2}=\hat{\lambda}{}^{2}\acute{x}{}^{2}$ by eliminating $\dot{x}\cdot\acute{x}$ from 
this pair of equations.  
Evidently, $\dot{x}{}^{2}=\hat{\lambda}{}^{2}\acute{x}{}^{2}$ is compatible with Eqs. (\ref{2.1}), 
if and only if 
\begin{align}
\hat{\lambda}=0 \,. 
\label{7.7}
\end{align}
In this way, the Lagrange multiplier field $\hat{\lambda}$ is necessarily determined to be 
zero.\footnote{Using Eq. (\ref{4.10}) considered for the restricted reparametrization (\ref{2.26}),  
we can show that the condition (\ref{7.7}) leads to 
$\hat{\lambda}^{\prime}:=\lambda^{\prime\;\! 1}/\lambda^{\prime\;\! 0} =0$.  
The condition (\ref{7.7}) is thus preserved under the restricted reparametrization. 
Correspondingly, the action $S_{6}$ with Eq. (\ref{7.7}), i.e., 
$$ 
\tilde{S}_{6}=\int_{\Xi} d^2 \xi \bigg[\:\!
\frac{i}{2} \Big(\bar{\mathscr{Z}}_{A} \dot{\mathscr{Z}}^{A} 
-\mathscr{Z}^{A} \dot{\bar{\mathscr{Z}}}_{A} \Big) 
+\varrho \!\:\bar{\mathscr{Z}}_{A} \mathscr{Z}^{A} \bigg] \,, 
$$
remains invariant under the restricted reparametrization (\ref{2.26}). 
As mentioned under Eq. (\ref{8.1}), the condition (\ref{7.7}) 
can be treated as a gauge-fixing condition.  
Then, the restricted reparametrization is considered to represent 
a residual gauge symmetry remaining in $\tilde{S}_{6}$. 
}    
As a result, the transformation rule (\ref{7.2}) and Eq. (\ref{7.3}) take the simple forms    
\begin{align}
\varrho \rightarrow \varrho^{\prime}=\varrho+\dot{\theta} \,, 
\label{7.8}
\\
\dot{\mathscr{Z}}^{A} -i\varrho \mathscr{Z}^{A}=0 \,.
\label{7.9}
\end{align}
Equation (\ref{7.9}) can immediately be solved as 
\begin{align}
\mathscr{Z}^{A} (\tau, \sigma)
=\exp\!\bigg[\!\: i \int_{0}^{\tau} \varrho(\tilde{\tau}, \sigma) d\tilde{\tau} \bigg] 
\mathscr{Z}^{A} (0, \sigma) \,, 
\label{7.10}
\end{align}
which is consistent with the local phase transformation 
$\mathscr{Z}^{A} \rightarrow \mathscr{Z}^{\prime A}=e^{i\theta} \mathscr{Z}^{A}$ 
supplemented by the transformation (\ref{7.8}), and with the set of Eqs. (\ref{7.4})--(\ref{7.6}). 
Because $\mathscr{Z}^{A}$ satisfies Eq. (\ref{7.5}), 
it is considered an element of 
the {\em null} twistor space 
$\mathbf{N}:=\big\{ \big(Z^{A} \big) \in \mathbf{T} \, \big|\, 
\bar{Z}_{A} Z^{A}=0 \big\}$. 
The solution (\ref{7.10}) demonstrates that the twistor 
$\mathscr{Z}^{A} (\tau, \sigma)$ for 
each $\sigma$ enjoys a circular motion in $\mathbf{N}$. 
However, because of  
$[ \mathscr{Z}^{A} (\tau, \sigma) ] =[ \mathscr{Z}^{A} (0, \sigma) ]$,  
this motion is not observed in the projective null twistor space 
$\mathbf{PN}:=\big\{ \big[ \big(Z^{A} \big) \big] \in \mathbf{PT} \, \big|\, 
\bar{Z}_{A} Z^{A}=0 \big\}$. 
In $\mathbf{PN}$, 
the projective twistors $[ \mathscr{Z}^{A} (\tau, \sigma) ]$ 
$(\sigma_{0}\leq \sigma \leq \sigma_{1})$ 
describe a (static) smooth curve parametrized by $\sigma$.

The canonical momenta conjugate to the twistors $\mathscr{Z}^{A}$ and 
$\bar{\mathscr{Z}}_{A}$ are found from the Lagrangian (\ref{7.1}) to be  
\begin{align}
\varPi_{A}:=\frac{\partial\mathcal{L}_{6}}{\partial \dot{\mathscr{Z}}^{A}}
=\frac{i}{2} \bar{\mathscr{Z}}_{A} \,, 
\qquad
\bar{\varPi}^{A}:=
\frac{\partial\mathcal{L}_{6}}{\partial \dot{\bar{\mathscr{Z}}}_{A}}
=-\frac{i}{2} \mathscr{Z}^{A} \,.
\label{7.11}
\end{align}
These equations are read as constraints in the phase space 
coordinatized by the canonical variables 
$\big(\mathscr{Z}^{A}, \bar{\mathscr{Z}}_{A}, \varPi_{A}, \bar{\varPi}^{A} \big)$. 
All the nonvanishing Poisson brackets between the canonical variables 
are included in  
\begin{subequations}
\label{7.12}
\begin{align}
\big\{ \mathscr{Z}^{A}(\tau, \sigma), \varPi_{B}(\tau, \tilde{\sigma}) \big\}_{\rm P}
&=\delta^{A}_{B} \delta(\sigma-\tilde{\sigma}) \,, 
\label{7.12a}
\\
\big\{ \bar{\mathscr{Z}}_{A}(\tau, \sigma), 
\bar{\varPi}^{B}(\tau, \tilde{\sigma}) \big\}_{\rm P}
&=\delta^{B}_{A} \delta(\sigma-\tilde{\sigma}) \,. 
\label{7.12b}
\end{align} 
\end{subequations}
With the relevant Poisson brackets, 
we see that the constraints in Eq. (\ref{7.11}) constitute  
second-class constraints. Then, in accordance with the Dirac formulation 
for constrained systems, the Dirac bracket $\{~,~\}_{\rm D}$ is defined in 
a manner such that  
\begin{subequations}
\label{7.13}
\begin{align}
\big\{ \mathscr{Z}^{A}(\tau, \sigma),  \bar{\mathscr{Z}}_{B}(\tau, \tilde{\sigma}) \big\}_{\rm D}
&=-i\delta^{A}_{B} \delta(\sigma-\tilde{\sigma}) \,, 
\label{7.13a}
\\
\big\{ \mathscr{Z}^{A}(\tau, \sigma),  \mathscr{Z}^{B}(\tau, \tilde{\sigma}) \big\}_{\rm D}
&=\big\{ \bar{\mathscr{Z}}_{A} (\tau, \sigma), 
\bar{\mathscr{Z}}_{B} (\tau, \tilde{\sigma}) \big\}_{\rm D}
=0 \,.
\label{7.13b}
\end{align}
\end{subequations}
Using the Dirac brackets (\ref{7.13}), we can readily verify that
\begin{subequations}
\label{7.14}
\begin{align}
\{ \mathcal{R}(\tau, \sigma), \mathcal{R}(\tau, \tilde{\sigma}) \}_{\rm D}
&=\big( \mathcal{R}(\tau, \sigma)+\mathcal{R}(\tau, \tilde{\sigma}) \big) 
\acute{\delta}(\sigma-\tilde{\sigma}) \,, 
\label{7.14a}
\\
\{ \mathcal{S}(\tau, \sigma), \mathcal{R}(\tau, \tilde{\sigma}) \}_{\rm D}
&=\mathcal{S}(\tau, \tilde{\sigma}) \acute{\delta}(\sigma-\tilde{\sigma}) \,, 
\label{7.14b}
\\
\{ \mathcal{S}(\tau, \sigma), \mathcal{S}(\tau, \tilde{\sigma}) \}_{\rm D} 
&=0 \,,
\label{7.14c}
\end{align}
\end{subequations}
where $\acute{\delta}(\sigma-\tilde{\sigma}):=
\frac{\partial}{\partial \sigma} \delta(\sigma-\tilde{\sigma})$. 
Hence, $\mathcal{R}$ and $\mathcal{S}$ constitute a closed algebra at 
the classical level, and therefore, there arise no further constraints associated 
with Eqs. (\ref{7.4}) and (\ref{7.5}). 
Equation (\ref{7.14a}) is precisely the (classical) Virasoro algebra 
without central extension written in a continuous-parameter form. 
This Virasoro algebra implies that $\mathcal{R}$ is a generator of  
the $\sigma$-reparametrization  
$\sigma  \rightarrow \sigma^{\prime}=\sigma^{\prime} (\sigma)$. 
In fact, the transformation rule of $\mathscr{Z}^{A}$ under the infinitesimal 
$\sigma$-reparametrization $\sigma^{\prime}=\sigma-\varepsilon(\sigma)$, i.e., 
\begin{align}
\delta \mathscr{Z}^{A}(\tau, \sigma)
&:=\mathscr{Z}^{\prime A}(\tau, \sigma)
-\mathscr{Z}^{A}(\tau, \sigma)
\nonumber 
\\
&=\varepsilon(\sigma) \acute{\mathscr{Z}}^{A}(\tau, \sigma)
+\frac{1}{2} \acute{\varepsilon}(\sigma) \mathscr{Z}^{A}(\tau, \sigma) \,, 
\label{7.15}
\end{align}
can be written as 
\begin{align}
\delta \mathscr{Z}^{A}(\tau, \sigma)
=-\int_{\sigma_{0}}^{\sigma_{1}}
 d \tilde{\sigma} \varepsilon(\tilde{\sigma}) \big\{ \mathcal{R}(\tau, \tilde{\sigma}), 
\mathscr{Z}^{A}(\tau, \sigma) \big\}_{\rm D} \,. 
\label{7.16}
\end{align}
Equations (\ref{7.14a}) and (\ref{7.14b}) immediately lead to the transformation rules 
of $\mathcal{R}$ and $\mathcal{S}$ under the infinitesimal 
$\sigma$-reparametrization.  
The canonical quantization of the twistor variables  
is carried out with the canonical commutation relations 
defined from the Dirac brackets (\ref{7.13}). 
In the quantization procedure, it is important to investigate whether 
the algebra (\ref{7.14}) is modified or not at the quantum-theoretical level. 
The study of quantization is in progress, and the details will be provided   
in the future.

Finally, we mention a resemblance between the twistor formulation of tensionless strings and 
that of massive particles. 
For this purpose, here, we consider a {\em closed} tensionless string characterized by 
the periodicity
\begin{align}
x^{\alpha\dot{\alpha}}(\tau, 2\pi)=x^{\alpha\dot{\alpha}}(\tau, 0) \,, 
\qquad 
\mathcal{P}_{\alpha\dot{\alpha}}(\tau, 2\pi)
=\mathcal{P}_{\alpha\dot{\alpha}}(\tau, 0) \,, 
\label{7.17}
\end{align}
where $\mathcal{P}_{\alpha\dot{\alpha}}$ is the (negative) canonical momentum 
\begin{align}
\mathcal{P}_{\alpha\dot{\alpha}}:= 
- \frac{\partial \mathcal{L}_{6}}{\partial \dot{x}^{\alpha\dot{\alpha}}} 
=\bar{\varpi}_{\alpha} \varpi_{\dot{\alpha}} \,.
\label{7.18}
\end{align}
The periodicity (\ref{7.17}) can be translated into the condition 
$\mathscr{Z}^{A}(\tau, 2\pi)=e^{i\gamma(\tau)}\mathscr{Z}^{A}(\tau, 0)$ 
with a phase $\gamma$, which generally depends on $\tau$. 
Regarding $\mathscr{Z}^{A}$ as a projective twistor,  we set $\gamma=1$, 
and thereby, the above condition for $\mathscr{Z}^{A}$ results in the periodicity 
\begin{align}
\mathscr{Z}^{A}(\tau, 2\pi)=\mathscr{Z}^{A}(\tau, 0) \,. 
\label{7.19}
\end{align}
From this, it follows that 
$\mathscr{Z}^{A}$ can be Fourier-expanded as 
$\mathscr{Z}^{A}(\sigma)=\sum_{n\in{\Bbb Z}} \mathscr{Z}^{A}_{n} e^{in\sigma}$, 
or in terms of the spinor representation 
$(\psi^{\alpha}, \varpi_{\dot{\alpha}})$  
with $\psi^{\alpha}:=ix^{\alpha\dot{\alpha}} \varpi_{\dot{\alpha}}\;\!$, as  
\begin{align}
\psi^{\alpha}(\sigma)
=\sum_{n\in{\Bbb Z}} \psi^{\alpha}_{n} e^{in\sigma} \,, 
\qquad 
\varpi_{\dot{\alpha}}(\sigma)
=\sum_{n\in{\Bbb Z}} \varpi_{\dot{\alpha} n} e^{in\sigma} \,. 
\label{7.20}
\end{align}
Here, the $\tau$-dependence in each Fourier expansion is to be understood.

The angular-momentum density of the tensionless string, 
$\mathcal{M}^{\alpha\beta\dot{\alpha}\dot{\beta}}:= 
x^{\alpha \dot{\alpha}} \mathcal{P}^{\beta\dot{\beta}}
-x^{\beta\dot{\beta}} \mathcal{P}^{\alpha \dot{\alpha}}$,  
can be written as 
\begin{align}
\mathcal{M}^{\alpha\beta\dot{\alpha}\dot{\beta}}= 
i\psi^{(\alpha} \bar{\varpi}^{\beta)} \epsilon^{\dot{\alpha}\dot{\beta}} 
-i\epsilon^{\alpha\beta} \bar{\psi}^{(\dot{\alpha}} \varpi^{\dot{\beta})}  \,.
\label{7.21}
\end{align}
From the 4-momentum and angular-momentum densities 
$\mathcal{P}_{\alpha\dot{\alpha}}$ and 
$\mathcal{M}^{\alpha\beta\dot{\alpha}\dot{\beta}}$, 
the 4-momentum and angular momentum of the closed tensionless string 
are defined by 
\begin{align}
\mathscr{P}_{\alpha\dot{\alpha}}=\int_{0}^{2\pi} \frac{d\sigma}{2\pi}
\mathcal{P}_{\alpha\dot{\alpha}}(\sigma) \,, 
\qquad 
\mathscr{M}^{\alpha\beta\dot{\alpha}\dot{\beta}}
=\int_{0}^{2\pi} \frac{d\sigma}{2\pi}
\mathcal{M}^{\alpha\beta\dot{\alpha}\dot{\beta}}(\sigma) \,.
\label{7.22}
\end{align}
In terms of the twistor Fourier-expansion coefficients 
\begin{align}
\mathscr{Z}^{A}_{n}=\big( \psi^{\alpha}_{n}, \varpi_{\dot{\alpha} n} \big) \,, 
\qquad 
\bar{\mathscr{Z}}_{An}=\big( \bar{\varpi}_{\alpha n},  \bar{\psi}^{\dot{\alpha}}_{n} \big) \,,
\label{7.23}
\end{align}
satisfying $\overline{\psi^{\alpha}_{n}}=\bar{\psi}^{\dot{\alpha}}_{-n}$ 
and $\overline{\varpi_{\dot{\alpha} n}}=\bar{\varpi}_{\alpha\, -n}$,  
the 4-momentum $\mathscr{P}_{\alpha\dot{\alpha}}$ and the angular momentum 
$\mathscr{M}^{\alpha\beta\dot{\alpha}\dot{\beta}}$ can be expressed as  
\begin{align}
\mathscr{P}_{\alpha\dot{\alpha}}&=\sum_{n\in{\Bbb Z}} 
\bar{\varpi}_{\alpha\, -n} \varpi_{\dot{\alpha} n} \,, 
\label{7.24}
\\
\mathscr{M}^{\alpha\beta\dot{\alpha}\dot{\beta}}&=\sum_{n\in{\Bbb Z}} 
\Big( i\psi^{(\alpha}_{n} \bar{\varpi}^{\beta)}_{-n} \epsilon^{\dot{\alpha}\dot{\beta}} 
-i\epsilon^{\alpha\beta} \bar{\psi}^{(\dot{\alpha}}_{-n} \varpi^{\dot{\beta})}_{n}  \Big)\,.
\label{7.25}
\end{align}
It is remarkable that Eqs. (\ref{7.24}) and (\ref{7.25}) are 
very similar to the expressions of 4-momentum and angular momentum 
of a massive particle written in spinorial forms\cite{Pen2, Per, Hug}. 
The only difference between the 4-momentum and angular momentum of  
the closed tensionless string and those of a massive particle is that the former, 
namely, $\mathscr{P}_{\alpha\dot{\alpha}}$ and  
$\mathscr{M}^{\alpha\beta\dot{\alpha}\dot{\beta}}$,  
are composed of an {\em infinite} number of constituent 2-component spinors. 
In twistor theory, a massive particle (or system) is formulated  
with a set of $N (\geq 2)$ twistors, and accordingly,    
the 4-momentum and angular momentum of this particle are written 
in terms of component spinors of the $N$ twistors. 
Analyzing invariance of these 4-momentum and angular momentum, 
we see that the massive particle possesses the global internal symmetry 
characterized by an inhomogeneous group that involves $U(N)$ 
as a homogeneous subgroup\cite{Per, Pen2, Hug}.  
Such an analysis may also be performed for 
$\mathscr{P}_{\alpha\dot{\alpha}}$ and  
$\mathscr{M}^{\alpha\beta\dot{\alpha}\dot{\beta}}\,$; 
then, it would be interesting to examine what internal symmetry is allowed in 
the system of a closed tensionless string   
without contradiction to the $\sigma$-reparametrization.

The squared mass of the closed tensionless string may be defined by 
$M^{2}=\mathscr{P}_{\alpha\dot{\alpha}} \mathscr{P}^{\alpha\dot{\alpha}}$ 
\cite{LRSS, GRR}. 
Substituting Eq. (\ref{7.24}) into this expression and using  
$\varpi_{\dot{\alpha} n} \varpi^{\dot{\alpha}}_{n}=0$, we have 
\begin{equation}
M^{2}=\sum_{\substack{m,\:\! n \in \Bbb{Z} \\
m\neq n}} 
\big| \varpi_{\dot{\alpha} m} \varpi^{\dot{\alpha}}_{n} \big|{}^{2} \,.
\label{7.26}
\end{equation}
Note here that only the products of different Fourier coefficients of 
$\varpi_{\dot{\alpha}}(\sigma)$ contribute to $M^{2}$.
Equation (\ref{7.26}) has the same form as the squared mass of a massive particle 
written in a spinorial form,    
except that Eq. (\ref{7.26}) is an infinite series. 
Details on $M^{2}$, including its quantum-theoretical properties, should be 
studied in the future.

\section{Summary and discussion}

We have provided and investigated spinor and twistor formulations of  
tensionless bosonic strings in 4-dimensional Minkowski space $\mathbf{M}$. 
We started from the first-order formalism defined by the action $\hat{S}_{1}$, 
which is equivalent to the Nambu-Goto action in the tensionful case and 
reduces to the Schild action in a particular gauge. 
In the tensionless case, $\hat{S}_{1}$ becomes the action $S_{1}$ yielding     
the two constraint equations (\ref{2.19}) and (\ref{2.20}). 
These equations were solved in terms of 2-component spinors, 
$\pi_{\dot{\alpha}}$ and $\bar{\pi}_{\alpha}$, 
and accordingly, $S_{1}$ was written as the spinorial action   
$S_{2}$ that governs the spinor formulation of tensionless strings. 
It was pointed out that in this formulation, the pair of the Virasoro constraints  
$P^{2}=0$ and $\acute{x}\cdot P=0$ takes a concise form (\ref{3.15}). 
To find an alternative spinorial action convenient for our study, 
we also considered the action $S_{3}$ equivalent to $S_{2}$,  
although $S_{3}$ remains invariant only under the restricted reparametrization (\ref{4.2}).   
Then, with the aid of the vector field $\lambda^{j}$, 
the action $S_{3}$ was improved to the alternative spinorial action $S_{4}$ 
in such a way that the full reparametrization invariance at the action level is restored.   
Because $S_{4}$ reduces to $S_{2}$ under a certain condition,  
$S_{4}$ can be treated as a fundamental action governing the spinor formulation of 
tensionless strings.

Beginning with the action $S_{5}$, 
we constructed a genuine twistor formulation of tensionless bosonic strings  
in accordance with a concept in twistor theory 
that twistors are more primitive than the space-time coordinates. 
To describe tensionless strings in $\mathbf{M}$, 
the action $S_{5}$ was modified to $S_{6}$ so as to involve the null twistor condition (\ref{5.2}).   
It was proven that the modified action $S_{6}$ reduces to the spinorial action $S_{4}$,  
and consequently, $S_{6}$ was recognized as a twistorial action for the tensionless string.  
The action $S_{6}$ was also obtained systematically by applying the gauge principle 
to the global phase symmetry inherent in $S_{5}$. 
Examining the local internal symmetries of the Lagrangian (\ref{5.8}), we made it 
clear that $S_{6}$ is defined for the projective twistor $[Z^{A}]$ rather than 
the (nonzero) twistor $Z^{A}$.

Classical analyses in the twistor formulation were carried out on the basis of the 
action $S_{6}$ with the Lagrangian (\ref{7.1}). 
It was shown through an analysis using the condition (\ref{2.1}) that 
the Lagrange multiplier field $\hat{\lambda}$  
eventually vanishes. This implies that $\hat{\lambda}=0$ is necessary  
so that $S_{6}$ can certainly be an action for the tensionless string. 
In order to incorporate the condition $\hat{\lambda}=0$ into the twistorial expressions  
without referring to the analysis carried out in terms of the space-time coordinates, 
we need to add the new term  
\begin{align}
S_{\beta\hat{\lambda}}=\int_{\varXi} d^2 \xi  \:\!\beta \hat{\lambda} 
\label{8.1}
\end{align}
to $S_{6}$. Here, $\beta$ is a new Lagrange multiplier field on $\varXi$. 
Varying $\beta$ in $\hat{S}_{6}:=S_{6}+S_{\beta\hat{\lambda}}$ yields 
the condition $\hat{\lambda}=0$, and therefore $\hat{S}_{6}$ can be considered  
a {\em complete} action governing the twistor formulation of tensionless strings. 
The additional term $S_{\beta\hat{\lambda}}$ can be found in  
a gauge-fixing procedure for the gauge symmetries inherent in $S_{6}$. 
Details on the gauge-fixing procedure will be reported in a forthcoming paper.

By putting $\hat{\lambda}=0$ in $S_{6}$, it becomes the action 
that has the form of $\sigma$-integral of the action (\ref{B.4}) given in Appendix B. 
Since the action (\ref{B.4}) describes a massless spinless particle, 
$S_{6}$ with $\hat{\lambda}=0$ turns out to describe a set of massless spinless particles  
arranged along a spacelike curve in $\mathbf{M}$.  
Here, the condition (\ref{7.4}) guarantees that this set of particles moves collectively  
in a direction perpendicular to the curve. 
The tensionless string is thus regarded as a string that consists of massless spinless particles 
subjected to the constraint (\ref{7.4}). 
This result is consistent with Eqs. (\ref{2.22}) and (\ref{2.23}). 
Recalling that the world line of a massless spinless particle,  
which is merely a light ray in $\mathbf{M}$, is represented by a (static) point in 
$\mathbf{PN}$,  we see that the world sheet of a tensionless string is represented by 
a (static) smooth curve in $\mathbf{PN}$. 
This is consistent with the fact mentioned under Eq. (\ref{7.10}). 
Comparing Eq. (\ref{7.5}) with Eq. (\ref{B.9}) makes it evident that 
the constituent massless particles of the tensionless string are spinless. 
If a tensionless string consisting of massless spinning bosonic particles,  
which we refer to here as the tensionless spinning bosonic string, is well defined, 
then the condition $\mathcal{S}=s$ is probably obtained instead of Eq. (\ref{7.5}). 
It may seem that the modification to the tensionless spinning bosonic string can easily be  
accomplished by adding $S_{\varrho}:=-2s \int_{\varXi} d^2 \xi \varrho$ to $S_{6}$.   
However, unlike the twistor formulation of massless particles provided in Appendix B,  
this is not allowed, because $S_{\varrho}$ remains invariant neither  
under the full reparametrization nor under the gauge transformation. 
Adding $S_{\varrho}$ to $S_{6}$ explicitly spoils the invariance properties of $S_{6}$.  
To avoid this trouble, we need to seek a satisfactory way of incorporating  
the condition $\mathcal{S}=s$ to the current formulation.

Studying the tensionless spinning bosonic string expressed in twistorial terms  
would lead to a supertwistor formulation of tensionless superstrings in 4 dimensions.  
Also, our twistor formulation of tensionless strings is perhaps  
related to the twistor string theory proposed by Witten \cite{Wit}. 
We hope to make these points clear in the near future.

\appendix

\section{Solving the Simultaneous Equations (\ref{2.19}) and (\ref{2.20})}

This appendix is devoted to solving the simultaneous equations (\ref{2.19}) and (\ref{2.20}) 
in terms of 2-component spinors. In the beginning,   
we briefly summarize the 2-component spinor notation and related conventions \cite{PM, HT, PR1}.

The complex conjugate of a 2-component contravariant spinor $\phi^{\alpha}$ ($\alpha=0, 1$) 
is denoted as 
$\bar{\phi}{}^{\dot{\alpha}}:=\overline{\phi^{\alpha}}$  ($\dot{\alpha}=\dot{0}, \dot{1}$). 
The corresponding covariant spinors $\phi_{\alpha}$ and $\bar{\phi}_{\dot{\alpha}}$ are given by  
\begin{align}
\phi_{\alpha}=\phi^{\beta}\epsilon_{\beta\alpha}\,, 
\qquad 
\bar{\phi}_{\dot{\alpha}}=\bar{\phi}{}^{\dot{\beta}}\epsilon_{\dot{\beta}\dot{\alpha}}\,, 
\qquad
\phi^{\alpha}=\epsilon^{\alpha\beta} \phi_{\beta} \,, 
\qquad
\bar{\phi}{}^{\dot{\alpha}}=\epsilon^{\dot{\alpha}\dot{\beta}} \bar{\phi}_{\dot{\beta}} \,, 
\label{A1} 
\end{align}
where the rank-2 $\epsilon$-spinors are defined by 
\begin{align}
\epsilon_{\alpha\beta}=\epsilon_{\dot{\alpha}\dot{\beta}}
=\epsilon^{\alpha\beta}=\epsilon^{\dot{\alpha}\dot{\beta}}
= \left(
\begin{array}{cc}
0  & 1 \\
-1 & 0 
\end{array}
\right) . 
\label{A2}
\end{align}
It is obvious that 
$\epsilon_{\alpha\gamma} \epsilon^{\beta\gamma}=\delta_{\alpha}{}^{\beta}$ 
and $\epsilon_{\dot{\alpha}\dot{\gamma}} \epsilon^{\dot{\beta}\dot{\gamma}}
=\delta_{\dot{\alpha}}{}^{\dot{\beta}}$.

Next, we introduce the sigma matrices 
\begin{align}
\big(\sigma_{\mu}{}^{\alpha\dot{\alpha}} \big):=
\frac{1}{\sqrt{2}} (\sigma_{0}, \sigma_{1}, -\sigma_{2}, \sigma_{3}) \,, 
\qquad 
\big(\sigma^{\mu}{}_{\alpha\dot{\alpha}} \big):=
\frac{1}{\sqrt{2}} (\sigma_{0}, \sigma_{1}, \sigma_{2}, \sigma_{3}) \,, 
\label{A3}
\end{align}
where $\sigma_{0}$ denotes the $2\times2$ unit matrix, and 
$\sigma_{i}$ ($i=1, 2, 3$) denotes the usual Pauli matrices. 
The matrix entries $\sigma_{\mu}{}^{\alpha\dot{\alpha}}$ and 
$\sigma^{\mu}{}_{\alpha\dot{\alpha}}$ are related by 
\begin{align}
\sigma^{\mu}{}_{\alpha\dot{\alpha}}
=\eta^{\mu\nu} \sigma_{\nu}{}^{\beta\dot{\beta}} 
\epsilon_{\beta\alpha} \epsilon_{\dot{\beta} \dot{\alpha}} \,, 
\qquad 
\sigma_{\mu}{}^{\alpha\dot{\alpha}}
=\eta_{\mu\nu} \epsilon^{\alpha\beta} \epsilon^{\dot{\alpha} \dot{\beta}} 
\sigma^{\nu}{}_{\beta\dot{\beta}}  \,, 
\label{A4}
\end{align}
and satisfy 
\begin{align}
\sigma^{\mu}{}_{\alpha\dot{\alpha}} \sigma_{\mu}{}^{\beta\dot{\beta}} 
&=\delta_{\alpha}{}^{\beta} \delta_{\dot{\alpha}}{}^{\dot{\beta}} \,, 
\label{A5}
\\
\sigma^{\mu}{}_{\alpha\dot{\alpha}} \sigma_{\nu}{}^{\alpha\dot{\alpha}} 
&=\delta^{\mu}{}_{\nu} \,. 
\label{A6}
\end{align}
Any vectors $V^{\mu}$ and $V_{\mu}$ are expressed  
in the 2-component spinor notation as 
\begin{align}
V^{\alpha\dot{\alpha}}=V^{\mu} \sigma_{\mu}{}^{\alpha\dot{\alpha}} \,,
\qquad 
V_{\alpha\dot{\alpha}}=V_{\mu} \sigma^{\mu}{}_{\alpha\dot{\alpha}} \,,
\label{A7}
\end{align}
which can be inversely solved using Eq. (\ref{A6}): 
\begin{align}
V^{\mu}=\sigma^{\mu}{}_{\alpha\dot{\alpha}} V^{\alpha\dot{\alpha}} \,, 
\qquad 
V_{\mu}=\sigma_{\mu}{}^{\alpha\dot{\alpha}} V_{\alpha\dot{\alpha}} \,. 
\label{A8}
\end{align}
By using Eq. (\ref{A4}), the inverse metric 
$\eta^{\alpha\beta\dot{\alpha}\dot{\beta}}
=\eta^{\mu\nu} \sigma_{\mu}{}^{\alpha\dot{\alpha}} \sigma_{\nu}{}^{\beta\dot{\beta}}$ 
and the metric 
$\eta_{\alpha\beta\dot{\alpha}\dot{\beta}}
=\eta_{\mu\nu} \sigma^{\mu}{}_{\alpha\dot{\alpha}} \sigma^{\nu}{}_{\beta\dot{\beta}}$  
can be written as 
\begin{align}
\eta^{\alpha\beta\dot{\alpha}\dot{\beta}}
=\epsilon^{\alpha\beta} \epsilon^{\dot{\alpha} \dot{\beta}} \,,
\qquad 
\eta_{\alpha\beta\dot{\alpha}\dot{\beta}}
=\epsilon_{\alpha\beta} \epsilon_{\dot{\alpha} \dot{\beta}} \,. 
\label{A9}
\end{align}
Similarly, 
the Levi-Civita symbol $\epsilon^{\mu\nu\rho\sigma}$ $\big(\epsilon^{0123}=-1\big)$ can be 
written in the 2-component spinor notation as 
\begin{align}
\epsilon^{\alpha\beta\gamma\delta\dot{\alpha}\dot{\beta}\dot{\gamma}\dot{\delta}}
&=\epsilon^{\mu\nu\rho\sigma}
\sigma_{\mu}{}^{\alpha\dot{\alpha}} \sigma_{\nu}{}^{\beta\dot{\beta}} 
\sigma_{\rho}{}^{\gamma\dot{\gamma}} \sigma_{\sigma}{}^{\delta\dot{\delta}} 
\nonumber 
\\
&=i \Big( \epsilon^{\alpha\gamma} \epsilon^{\beta\delta} \epsilon^{\dot{\alpha}\dot{\delta}} 
\epsilon^{\dot{\beta}\dot{\gamma}} 
-\epsilon^{\alpha\delta} \epsilon^{\beta\gamma} \epsilon^{\dot{\alpha}\dot{\gamma}} 
\epsilon^{\dot{\beta}\dot{\delta}} \Big)\,. 
\label{A10}
\end{align}
Also, it is known that any real antisymmetric tensor of rank 2, $p_{\mu\nu}(=-p_{\nu\mu}$),  
can be expressed as  
\begin{align}
p_{\alpha\beta\dot{\alpha}\dot{\beta}}  
&=p_{\mu\nu} \sigma^{\mu}{}_{\alpha\dot{\alpha}} \sigma^{\nu}{}_{\beta\dot{\beta}} 
\nonumber 
\\
&=\psi_{\alpha\beta} \epsilon_{\dot{\alpha}\dot{\beta}}
+\bar{\psi}_{\dot{\alpha}\dot{\beta}} \epsilon_{\alpha\beta} \,, 
\label{A11}
\end{align}
where $\psi_{\alpha\beta}$ and its complex conjugate 
$\bar{\psi}_{\dot{\alpha}\dot{\beta}}$ are symmetric spinors defined by 
\begin{align}
\psi_{\alpha\beta}:=\frac{1}{4} \big( p_{\alpha\beta\dot{\gamma}}{}^{\dot{\gamma}}  
+p_{\beta\alpha\dot{\gamma}}{}^{\dot{\gamma}} \big) \,, 
\qquad 
\bar{\psi}_{\dot{\alpha}\dot{\beta}}:
=\frac{1}{4} \big( \bar{p}_{\gamma}{}^{\gamma}{}_{\dot{\alpha}\dot{\beta}}  
+\bar{p}_{\gamma}{}^{\gamma}{}_{\dot{\beta}\dot{\alpha}} \big) \,.
\label{A12}
\end{align}
Substituting Eqs. (\ref{A10}) and (\ref{A11}) into the spinor form 
of the Hodge dual tensor 
$\tilde{p}^{\mu\nu}:=\frac{1}{2} \epsilon^{\mu\nu\rho\sigma} p_{\rho\sigma}$, 
we have  
\begin{align}
\tilde{p}^{\alpha\beta\dot{\alpha}\dot{\beta}} 
&=\frac{1}{2} 
\epsilon^{\alpha\beta\gamma\delta\dot{\alpha}\dot{\beta}\dot{\gamma}\dot{\delta}}
p_{\gamma\delta\dot{\gamma}\dot{\delta}} 
\nonumber
\\
&=-i \psi^{\alpha\beta} \epsilon^{\dot{\alpha}\dot{\beta}}
+i \bar{\psi}^{\dot{\alpha}\dot{\beta}} \epsilon^{\alpha\beta} \,.
\label{A13}
\end{align}

Now, let us consider the simultaneous equations (\ref{2.19}) and (\ref{2.20}), 
or equivalently, 
\begin{align}
p_{\alpha\beta\dot{\alpha}\dot{\beta}}\;\!  p^{\alpha\beta\dot{\alpha}\dot{\beta}} &=0 \,, 
\label{A14}
\\
p_{\alpha\beta\dot{\alpha}\dot{\beta}}\;\!  \tilde{p}^{\alpha\beta\dot{\alpha}\dot{\beta}} &=0 \,. 
\label{A15}
\end{align}
From Eqs. (\ref{A11}) and (\ref{A13}), it follows that  
\begin{align}
p_{\alpha\beta\dot{\alpha}\dot{\beta}}\;\!  p^{\alpha\beta\dot{\alpha}\dot{\beta}}
&=2\Big( \psi_{\alpha\beta} \psi^{\alpha\beta}
+\bar{\psi}{}_{\dot{\alpha}\dot{\beta}} \bar{\psi}{}^{\dot{\alpha}\dot{\beta}} \Big) \,,
\label{A16}
\\
p_{\alpha\beta\dot{\alpha}\dot{\beta}}\;\!  \tilde{p}^{\alpha\beta\dot{\alpha}\dot{\beta}}
&=-2i \Big( \psi_{\alpha\beta} \psi^{\alpha\beta}
-\bar{\psi}_{\dot{\alpha}\dot{\beta}} \bar{\psi}^{\dot{\alpha}\dot{\beta}} \Big) \,. 
\label{A17}
\end{align}
Then, it is obvious that the pair of Eqs. (\ref{A14}) and (\ref{A15}) is equivalent to 
the pair of equations $\psi_{\alpha\beta} \psi^{\alpha\beta}=0$ and 
$\bar{\psi}_{\dot{\alpha}\dot{\beta}} \bar{\psi}^{\dot{\alpha}\dot{\beta}}=0$. 
Because the latter equation is merely the complex conjugate of the former equation, 
we can conclude that the pair of Eqs. (\ref{A14}) and (\ref{A15}) is equivalent to  
\begin{align}
\psi_{\alpha\beta} \psi^{\alpha\beta}=0 \,.
\label{A18}
\end{align}
Recall here the proposition that 
{\em any totally symmetric spinor can be uniquely 
decomposed into a symmetrized product of rank 1 spinors} \cite{HT, PR1}. 
For the symmetric spinor $\psi_{\alpha\beta}$, this proposition reads 
\begin{align}
\psi_{\alpha\beta}=\phi_{\alpha}\varphi_{\beta}+\phi_{\beta}\varphi_{\alpha} \,, 
\label{A19}
\end{align}
with rank 1 spinors $\phi_{\alpha}$ and $\varphi_{\alpha}$. 
Substituting Eq. (\ref{A19}) into Eq. (\ref{A18}) leads to $\phi_{\alpha} \varphi^{\alpha}=0$, 
which implies that $\varphi_{\alpha}=k \phi_{\alpha}$ $(k \in\Bbb{C})$. 
Hence, Eq. (\ref{A19}) reduces to  
$\psi_{\alpha\beta}=\bar{\pi}_{\alpha} \bar{\pi}_{\beta}$, 
where $\bar{\pi}_{\alpha}:=\sqrt{2k}\, \phi_{\alpha}$. 
(Here, following a convention in twistor theory, 
we put a ^^ ^^ bar'' over the undotted spinor, so that  
$\pi_{\dot{\alpha}}=\overline{\bar{\pi}_{\alpha}}\:\!$.)  
Substituting $\psi_{\alpha\beta}=\bar{\pi}_{\alpha} \bar{\pi}_{\beta}$ 
into Eq. (\ref{A11}), we have 
\begin{align}
p_{\alpha\beta\dot{\alpha}\dot{\beta}}  
=\bar{\pi}_{\alpha} \bar{\pi}_{\beta} \epsilon_{\dot{\alpha}\dot{\beta}}
+\pi_{\dot{\alpha}} \pi_{\dot{\beta}} \epsilon_{\alpha\beta} \,. 
\label{A20}
\end{align}
In this way, 
the simultaneous equations (\ref{A14}) and (\ref{A15}) are solved as Eq. (\ref{A20}) 
in terms of the spinors $\bar{\pi}_{\alpha}$ and $\pi_{\dot{\alpha}}$.  
We can readily verify by direct substitution that Eq. (\ref{A20})  
is a solution of Eqs. (\ref{A14}) and (\ref{A15}). 
Equation (\ref{A20}) is thus equivalent to the pair of Eqs. (\ref{A14}) and (\ref{A15}).

\section{A Twistor Formulation of Massless Particles}

In this appendix, we consider a twistor formulation of massless particles 
in connection with the twistor formulation of tensionless strings studied   
in the present paper. 
For a massless particle propagating in Minkowski space $\mathbf{M}$,  
the action corresponding to the action (\ref{6.4.1}) is given by 
\begin{align}
S_{\mathrm{mlp}}=\int_{\tau_0}^{\tau_1} d\tau 
\bigg[\;\! \frac{i}{2} \lambda  
\big(\bar{Z}_{A} DZ^{A} -Z^{A} \bar{D}\bar{Z}_{A} \big) \bigg] 
\label{B.1}
\end{align}
with $D:=\frac{d}{dt}-ia$. 
Here, $\lambda$, as well as $Z^{A}$ and $\bar{Z}_{A}$, is taken to be 
a scalar field on the 1-dimensional space  
$\mathcal{T}:=\{ (\tau)\:\! |\, \tau_{0} \leq \tau \leq \tau_{1} \}$, 
while $a$ is treated as a $U(1)$ gauge field on $\mathcal{T}$. 
(Hence, $a$ behaves as a real scalar-density field on $\mathcal{T}$.)  
Obviously, $S_{\mathrm{mlp}}$ is left invariant under the reparametrization 
$\tau \rightarrow \tau^{\prime} (\tau)$. 
In addition, it remains invariant under the complexified local scale transformation 
\begin{align}
Z^{A} &\rightarrow Z^{\prime A}=\upsilon(\tau) Z^{A} \,, 
\qquad 
\bar{Z}_{A} \rightarrow \bar{Z}^{\prime}_{A}
=\bar{\upsilon}(\tau) \bar{Z}_{A} \,, 
\label{B.2}
\end{align}
supplemented by the transformations  
\begin{align}
\lambda \rightarrow \lambda=|\upsilon|^{-2} \lambda \,, 
\qquad
a \rightarrow a^{\prime}=a+\dot{\theta}  
\label{B.3}
\end{align}
with $\theta:={1\over2}i \ln (\bar{\upsilon}/ \upsilon)$. 
Because of this scale invariance, 
the twistor $Z^{A}$ in Eq. (\ref{B.1}) is regarded as a projective twistor.

Assuming that $\lambda >0$, we can write $S_{\mathrm{mlp}}$ as 
\begin{align}
S_{\mathrm{mlp}}=\int_{\tau_0}^{\tau_1} d\tau 
\bigg[\;\! 
\frac{i}{2} \Big(\bar{\mathscr{Z}}_{A} \dot{\mathscr{Z}}^{A} 
-\mathscr{Z}^{A} \dot{\bar{\mathscr{Z}}}_{A} \Big) 
+a \!\:\bar{\mathscr{Z}}_{A} \mathscr{Z}^{A} \bigg] \,, 
\label{B.4}
\end{align}
where 
$\mathscr{Z}^{A}:=\sqrt{\lambda}\:\! Z^{A}$ and 
$\bar{\mathscr{Z}}_{A}:=\sqrt{\lambda}\:\! \bar{Z}_{A}$.   
In this expression, the complexified scale invariance of $S_{\mathrm{mlp}}$ 
is realized as the invariance under the local phase transformation 
\begin{align}
\mathscr{Z}^{A} \rightarrow \mathscr{Z}^{\prime A}=e^{i\theta} \mathscr{Z}^{A} \,, 
\qquad 
\bar{\mathscr{Z}}_{A} \rightarrow \bar{\mathscr{Z}}^{\prime}_{A}
=e^{-i\theta} \bar{\mathscr{Z}}_{A} \,, 
\label{B.5}
\end{align}
supplemented by the gauge transformation 
\begin{align}
a \rightarrow a^{\prime}=a+\dot{\theta}  \,. 
\label{B.6}
\end{align}

Unlike the twistor formulation of tensionless strings, 
now there is room to add the 1-dimensional Chern-Simons (CS) term 
\begin{align}
S_{\mathrm{1CS}}=-2s \int_{\tau_0}^{\tau_1} d\tau a \,,
\label{B.7}
\end{align}
with a real constant $s$,  
to $S_{\mathrm{mlp}}$ without spoiling the invariance properties of $S_{\mathrm{mlp}}$ 
stated above.  
In fact, $S_{\mathrm{1CS}}$ has the same invariance properties as $S_{\mathrm{mlp}}\,$; 
that is, $S_{\mathrm{1CS}}$ is left invariant both under the reparametrization 
$\tau \rightarrow \tau^{\prime} (\tau)$ and the gauge transformation (\ref{B.6}),  
provided $\theta$ satisfies an appropriate boundary condition 
such as $\theta(\tau_{1})=\theta(\tau_{0})$. 
Varying $a$ in the total action 
\begin{align}
\hat{S}_{\mathrm{mlp}}&=S_{\mathrm{mlp}} +S_{\mathrm{1CS}} 
\nonumber 
\\
& =\int_{\tau_0}^{\tau_1} d\tau 
\bigg[\;\! 
\frac{i}{2} \Big(\bar{\mathscr{Z}}_{A} \dot{\mathscr{Z}}^{A} 
-\mathscr{Z}^{A} \dot{\bar{\mathscr{Z}}}_{A} \Big) 
+a \big( \bar{\mathscr{Z}}_{A} \mathscr{Z}^{A} 
-2s \big) \bigg] \,, 
\label{B.8}
\end{align}
we have 
\begin{align}
{1\over 2} \bar{\mathscr{Z}}_{A} \mathscr{Z}^{A} =s \,.
\label{B.9}
\end{align}
This is precisely the helicity condition for a massless particle familiar in  
twistor theory \cite{PM, Pen2, PR, Pen}. 
Thus, we see that the action $\hat{S}_{\mathrm{mlp}}$ describes  
a massless spinning particle as well as a massless spinless particle. 
After the canonical quantization of the twistor variables is performed,  
the Weyl-ordered form of Eq. (\ref{B.9}) applied to an eigenfunction  
is read as the eigenvalue equation for the helicity operator, 
in which $s$ is understood as a helicity eigenvalue. 
By imposing the single-valuedness on the eigenfunction, 
it turns out that the allowed values of $s$ are restricted to 
integer and half-integer values in units such that $\hbar=1$.

Because $\hat{S}_{\mathrm{mlp}}$ involves the helicity condition at the action level, 
it is suitable for describing a massless particle with a fixed value of the helicity $s$. 
The action $\hat{S}_{\mathrm{mlp}}$ can be derived by gauging the phase symmetry 
of the action found by Shirafuji \cite{Shi}, and for this reason,  
$\hat{S}_{\mathrm{mlp}}$ may be referred to as the {\em gauged} Shirafuji action. 
This action has also been proposed by Bars and Pic\'{o}n  
in a somewhat different context \cite{BP}.

\end{document}